# Measurements and computational analysis on the natural decay of $^{176}$Lu

F. G. A. Quarati,[1,2] G. Bollen,[3,4,5] P. Dorenbos,[1] M. Eibach,[3,6] K. Gulyuz,[7] A. Hamaker,[3,5] C. Izzo,[3,5] D. K. Keblbeck,[7] X. Mougeot,[8] D. Puentes,[3,5] M. Redshaw,[3,7] R. Ringle,[3] R. Sandler,[7] J. Surbrook,[3,5] and I. Yandow[3,5]

[1]*Delft University of Technology, Faculty of Applied Science, Radiation Science and Technology Department, Mekelweg 15, 2629JB Delft, The Netherlands*

[2]*Gonitec BV, Johannes Bildersstraat 43, 2596EE Den Haag, The Netherlands*

[3]*National Superconducting Cyclotron Laboratory, East Lansing, Michigan, 48824, USA*

[4]*Facility for Rare Isotope Beams, East Lansing, Michigan, 48824, USA*

[5]*Department of Physics and Astronomy, Michigan State University, East Lansing, Michigan 48824, USA*

[6]*Institut fur Physik, Universitat Greifswald, 17487 Greifswald, Germany*

[7]*Department of Physics, Central Michigan University, Mount Pleasant, Michigan, 48859, USA*

[8]*Universite Paris-Saclay, CEA, List, Laboratoire National Henri Becquerel (LNE-LNHB), F-91120 Palaiseau, France*

**Background:** Mainly because of its long half-life and despite its scientific relevance, spectroscopic measurements of $^{176}$Lu forbidden $\beta$-decays are very limited and lack formulation of shape factors. A direct precise measurement of its Q value is also presently unreported. In addition, the description of forbidden decays provides interesting challenges for nuclear theory. The comparison of precise experimental results with theoretical calculations for these decays can help to test underlying models and can aid the interpretation of data from other experiments.

**Purpose:** Perform the first precision measurements of $^{176}$Lu $\beta$-decay spectra and attempt the observation of its electron capture decays, as well as perform the first precision direct measurement of the $^{176}$Lu $\beta$-decay Q value. Compare the shape of the precisely determined experimental $\beta$-spectra to theoretical calculations, and compare the end point energy to that obtained from an independent Q value measurement.

**Method:** The $^{176}$Lu $\beta$-decay spectra measurements and the search for electron capture decays were performed with an experimental set-up that employed lutetium-containing scintillator crystals and a NaI(Tl) spectrometer for coincidence counting. The $\beta$-decay Q value was determined via high-precision Penning trap mass spectrometry (PTMS) with the LEBIT facility at the National Superconducting Cyclotron Laboratory. The $\beta$-spectrum calculations were performed within the Fermi theory formalism with nuclear structure effects calculated using a shell model approach.

**Results:** Both $\beta$-transitions of $^{176}$Lu were experimentally observed and corresponding shape factors formulated in their entire energy ranges. The search for electron capture decay branches led to an experimental upper limit of $6.3\times10^{-6}$ relative to its $\beta$-decays. The $^{176}$Lu $\beta$-decay and electron capture Q values were measured using PTMS to be 1193.0(6) keV and 108.9(8) keV, respectively. This enabled precise $\beta$ end point energies of 596.2(6) keV and 195.3(6) keV to be determined for the primary and secondary $\beta$-decays, respectively. The conserved vector current hypothesis was applied to calculate the relativistic vector matrix elements. The $\beta$-spectrum shape was shown to significantly depend on the Coulomb displacement energy and on the value of the axial vector coupling constant $g_A$, which was extracted according to different assumptions.

**Conclusion:** The implemented self-scintillation method has provided unmatched observations of $^{176}$Lu, independently validated by the first direct measurements of its $\beta$-decay Q value by Penning trap mass spectrometry. Theoretical study of the main $\beta$-transition led to the extraction of very different effective $g_A$ and log*f* values, showing that a high-precision description of this transition would require a realistic nuclear structure with nucleus deformation.

## I. INTRODUCTION

Scientific advances that have strong impact, from fundamental physics to applications in everyday life, rely on accurate nuclear data [1]. The primordial, long-lived, $^{176}$Lu radionuclide possesses unique relevance in nuclear astrophysics, nucleosynthesis, geo- and extra-terrestrialchronology [2–8] and additional relevance in nuclear structure science and applications [9–12].

The dominant transition of $^{176}$Lu (see Fig. 1) is a forbidden β-transition and the study and description of such types of β-decay has itself shown to provide both experimental and theoretical challenges along with more practical applications, as outlined in [13]. For instance, precise measurements and theoretical descriptions of $\beta$-spectra could provide a method to extract a value for the weak axial vector coupling constant, $g_A$, as recently described by Haaranen *et al.* [14]. Besides, these kinds of $\beta$-decays can create significant, sometimes dominant background events in highly sensitive experiments, such as (neutrinoless) double $\beta$-decay, dark matter searches, and solar- and geo-neutrino experiments [15–17]. More specifically for $^{176}$Lu, recent developments have shown that precise knowledge of the $\beta$-spectrum shape is fundamental for a precise half-life determination by liquid scintillation counting [18–20]. Therefore, the availability of precise shape factors for $^{176}$Lu can enable a more accurate determination of its half-life, which is crucial for application of the Lu-Hf dating system in geochronology.

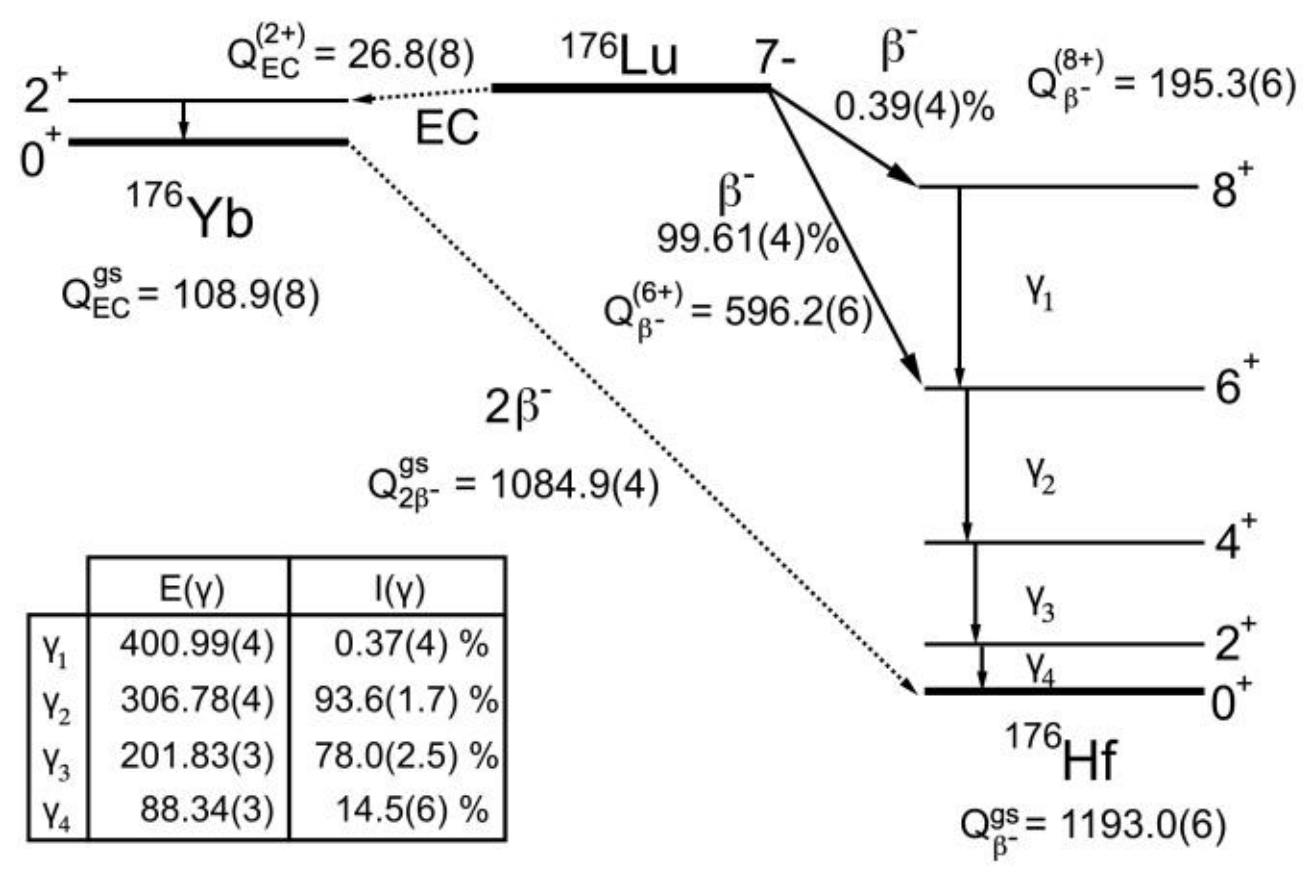

Figure 1. Decay scheme for Lu-Hf-Yb $A$ = 176 triplet. Solid arrows indicate the observed 1st forbidden non-unique $\beta$decays of $^{176}$Lu, and dotted arrows indicate the energetically allowed, but as yet unobserved 5th forbidden non-unique EC decay of $^{176}$Lu, and the double $\beta$-decay of $^{176}$Yb. The table inset shows the energies and absolute $\gamma$-ray intensities of the different transitions in $^{176}$Hf. All energies are in keV and Q values are from this work.

Early investigations on the natural radioactivity of lutetium include the work of Heyden and Wefelmeier, and that of Libby, in 1938 and 1939 respectively [21, 22]. Despite the numerous successive publications, the work of Dixon *et al.* in 1954 [23] and Prodi *et al.* in 1969 [24] remain the only available references for experimental $\beta$ end point and $\beta$-spectrum shape for the dominant (>99%) $\beta$-transition of $^{176}$Lu. Both of them, however, made evaluations of the end point energy that are inconsistent with recent atomic mass evaluation data (AME2020) [25], and both observed $\beta$-spectra that are in good agreement with the spectrum for an allowed transition. Therefore, experimental $\beta$-shape factors for $^{176}$Lu have been, until now, unavailable in the literature. Similarly, precise, direct measurements of the Q value of $^{176}$Lu have not been previously reported; an experimental Q value was previously obtained by Ketelaer *et al.* [26] from individual Penning trap measurements of parent and daughter atomic masses, but with an uncertainty of ~ 11 keV. Hence, a new direct measurement is called for.

As shown in Fig. 1 $^{176}$Lu decays via two 1st forbidden $\beta$-decay branches and with a half-life of 3.76(7)×$10^{10}$ y; foremost with a predominant transition (99.61 % probability) to the $6^+$ state of $^{176}$Hf and then to its ground state via de-excitation cascade ($4^+$, $2^+$ and $0^+$); secondly via a much weaker branch (0.39 % probability) to the $8^+$ state of $^{176}$Hf and then again to its ground state via deexcitation cascade ($6^+$, $4^+$, $2^+$ and $0^+$) [27]. The inset of Fig. 1 reports the energies and $\gamma$ emission probabilities of the de-excitation cascades. For readiness, these energies have been approximated to the nearest keV in the text. Electron capture (EC) decay of $^{176}$Lu to $^{176}$Yb is also energetically allowed, but such a decay to the $2^+$ state of $^{176}$Yb is 5th forbidden and modern decay experiments and geochemical tests have shown no indication of this decay branch [28–30]. $^{176}$Lu is also the most strongly bound isotope in a nearly stable triplet, so double $\beta$-decay of $^{176}$Yb → $^{176}$Hf is also energetically allowed. Again, experimental searches for this decay have been performed, but it has not been observed [31].

We describe herewith our efforts to improve the knowledge of $^{176}$Lu decay, aiming at precise evaluations of the Q values, the $\beta$-spectrum shapes, and their end point energies. We also formulate much stricter limits for its EC branches. In addition, theoretical calculations of these $\beta$decays and extensive comparisons of the results with our measurements have been performed. The current limits on the EC decay branches of $^{176}$Lu have been revised as well.

The paper is organized as follows: in Sec. IIA to IIC, we describe in detail the precise measurement of the $\beta$spectrum and its end point energy for $^{176}$Lu decay using Lu-containing scintillators, including a brief reintroduction of the self-scintillation method. In Sec. IID, we report on the investigation of the EC decay of $^{176}$Lu. In Sec. III, we describe the first direct measurement of the $^{176}$Lu $\beta$-decay Q value via Penning trap mass spectrometry measurements of the mass ratio of parent and daughter ions. The theoretical calculations of $^{176}$Lu $\beta$decay and subsequent analyses are depicted in Sec. IV. Finally, we summarize and discuss the results, their significance and envisaged implications in the last section.

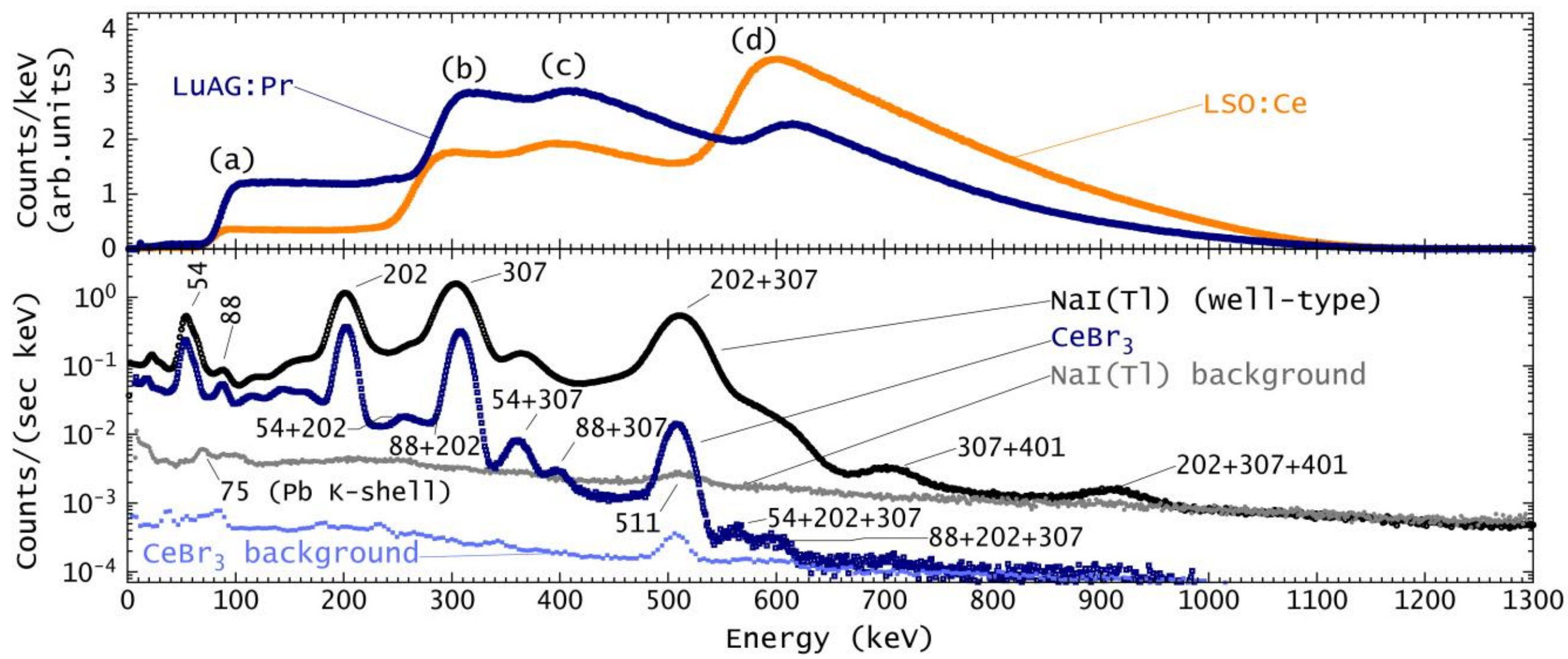

Figure 2. Top, self-scintillation spectra of LuAG:Pr and LSO:Ce, see text for further explanation. Bottom, $\gamma$-ray emission escaping LuAG:Pr and LSO:Ce and detected by NaI(Tl) and $CeBr_3$ spectrometer. All measurements are performed inside a lead castle. All reported values refer to the energy and are expressed in keV.

## II. $\beta$-SPECTRUM MEASUREMENTS

### A. The self-scintillation method for $^{176}$Lu

Spectroscopic $\beta$-decay measurements of long-lived nuclides such as $^{176}$Lu are challenging due to the low activity achieved by standard radioactive sources. A scintillator detector whose molecules contain the long-lived nuclide, either by natural occurrence or by means of doping, is particularly suitable to cope with this challenge. In this case, the radioactive source and the scintillator detector form a sort of calorimeter and, by means of pulse height spectroscopic measurements of the self-scintillation, alias intrinsic background, one can then detect the $\beta$-spectrum shape. Such self-scintillation methods were applied in the past and a notable example is the work of Beard and Kelly in 1961 [32]. More recently, we have applied the self-scintillation method to the measurement of the $\beta$-spectrum shape of $^{138}$La using $LaBr_3$:Ce and $CeBr_3$ scintillators [33–35].

In the past three decades, achievements in research and engineering have made available numerous Lu-containing scintillators such as LSO:Ce [1], LuAP:Ce [2], LYSO:Ce [3], LuAG:Pr4[4] and LuYAP:Pr[5] [36–40]. Because of their high density and fast response, these scintillators found their main applications in positron emission tomography. The natural occurrence of $^{176}$Lu generates in these scintillators a specific activity ranging from 150 Bq/cm$^3$ up to 300 Bq/cm$^3$, quite remarkable when compared to the over 30 billion year long half-life of $^{176}$Lu.

Current research on the physics of the scintillation mechanism includes relevant experiments and data on the so-called scintillation non-proportionality of the response (nPR). Namely, the number of scintillation photons produced by $\gamma$- and $\beta$-ray interactions, i.e. the so called scintillation light yield, is not linearly proportional to the energy deposited in the scintillator [41, 42]. Since the nPR particularly affects the energies below 100 keV, it is commonly expressed in percent of the scintillation light yield measured at higher energies, generally at 662 keV ($^{137}$Cs source).

Extensive and precise knowledge of nPR requires complex experiments involving, for instance, dedicated facilities for Compton coincidence technique [43] or highly monochromatic synchrotron radiation [44]. While nPR measurements with these techniques agree reasonably well, some unexplained discrepancy exists as reported by [45] in the case of $LaBr_3$:Ce and $LaCl_3$:Ce scintillators. Moreover, it was demonstrated that nPR can be affected by unintentional contamination [46] or even engineered by intentional co-doping [47, 48]. Therefore uncertainties in nPR can exist not only for a given scintillator compound but also among different samples of that compound. Still, using the experiments and studies of Payne et al. [49] and Khodyuk and Dorenbos [50, 51], we could assess and take into account the effects of nPR in the present work.

[1] LSO:Ce Lutetium Oxyorthosilicate doped with Cerium, $Lu_2SiO_5:Ce^{3+}$.

[2] LuAP:Ce Lutetium Aluminum Perovskite doped with Cerium, LuAlO3:Ce3+.

[3] LYSO:Ce Lutetium-Yttrium Oxyorthosilicate doped with Cerium, $Lu_{1.8}Y_{0.2}SiO_5:Ce^{3+}$.

[4] LuAG:Pr Lutetium Aluminum Garnet doped with Praseodymium, $Lu_3Al_5O_{12}:Pr^{3+}$.

[5] LuYAP:Pr Lutetium-Yttrium Aluminum Perovskite doped with Praseodymium, $(Lu_{0.75}Y_{0.25})_3Al_5O_{12}:Pr^{3+}$.

During the preparatory phases of our experiments, we considered and tested the Lu-containing scintillators mentioned above and concluded to focus our efforts on LSO:Ce and LuAG:Pr. The former, despite its strong nPR, is possibly the most common Lu-containing scintillator; the latter offers one of the most proportional responses and hence is least affected by nPR. Both of them have a wide commercial availability and have been well characterised in terms of nPR.

Given the relative complexity of the $\beta$-decay branch of $^{176}$Lu (Fig. 1) with two transitions and four excited levels, there is not a straight-forward method for its spectroscopic measurements, but rather various possible configurations. This aspect has been investigated using LSO:Ce and LuAG:Pr and is further addressed in the next subsection.

### B. Experimental apparatus and methods

Self-scintillation spectra measured with LSO:Ce and LuAG:Pr scintillators with sizes of 3.2 × 3.2 × 1 cm$^3$ (10.2 cm$^3$) and 8×8×8 mm$^3$ (0.5 cm$^3$), respectively, are reported in Fig. 2 (Top). The measurements were performed by coupling the scintillators to a photomultiplier tube (PMT) and placing the resulting assembly inside a low-activity lead-castle. Signal read out and acquisition involved widely available nuclear spectroscopy equipment such as a shaping amplifier and analog-to-digital converter of the NIM and CAMAC standard. Standard radioactive sources were used for channel to energy calibrations and evaluations of the energy resolution.

Both spectra are shaped by the dominant $\beta$-transition (99.6% probability) of $^{176}$Lu, and one can distinguish four main sawtooth-shaped peaks marked (a), (b), (c), and (d) in Fig. 2 (Top). With reference to the decay scheme of $^{176}$Lu (Fig. 1), these peaks can be identified with the detection of true sum coincidences of the $\beta$-particle with one or more of the three $^{176}$Hf de-excitation emissions of 88 keV, 202 keV and 307 keV. For example peak (a) is the detection of the $\beta$ plus the 88 keV emission. In this case the 202 keV and 307 keV escape the scintillator. For peak (b) only the 307 keV $\gamma$ escapes, for peak (c) only the 202 keV $\gamma$ escapes, and for peak (d) there is no escape and the $\beta$ plus all three final state levels are detected.

The spectra of Fig. 2 (Top) also include some events from the minor (0.4% probability) $\beta$-transition. If fully detected this should give rise to a fifth sawtooth-shape peak at 998 keV, which is not observed because of the large escape probability of the 401 keV even for the larger LSO:Ce scintillator. Therefore, most of the events related to the minor $\beta$-transition overlap with those of the dominant transition and cannot be distinguished. Events related to electron capture decay of $^{176}$Lu are also potentially present in the spectra of Fig. 2 (Top). However, we could not distinguish any related features. We report further on EC in the dedicated Section IID.

Another feature of both spectra in Fig. 2 (Top) is a much smaller sawtooth-shaped peak around 34 keV, which, for the LuAG:Pr scintillator, can be better observed in Fig. 4. This peak is related to K-shell x-ray emissions escaping the scintillator. These emissions originate from the internal conversion of mostly the 2$^+$ state of Hf and from the fluorescence of Lu. The binding energies of Hf and Lu are so close that their x-ray emissions overlap into a unique peak (which is also found in Fig. 2 (Bottom)) with an energy of about 54 keV; hence the position of the sawtooth-shaped peak in Fig. 2 (Top) at 88 – 54 = 34 keV. One more notable feature is the relatively more intense (d) peak observed with LSO:Ce. This is simply due to the much larger size of this scintillator and hence smaller escape probability for the 307 keV $\gamma$-ray.

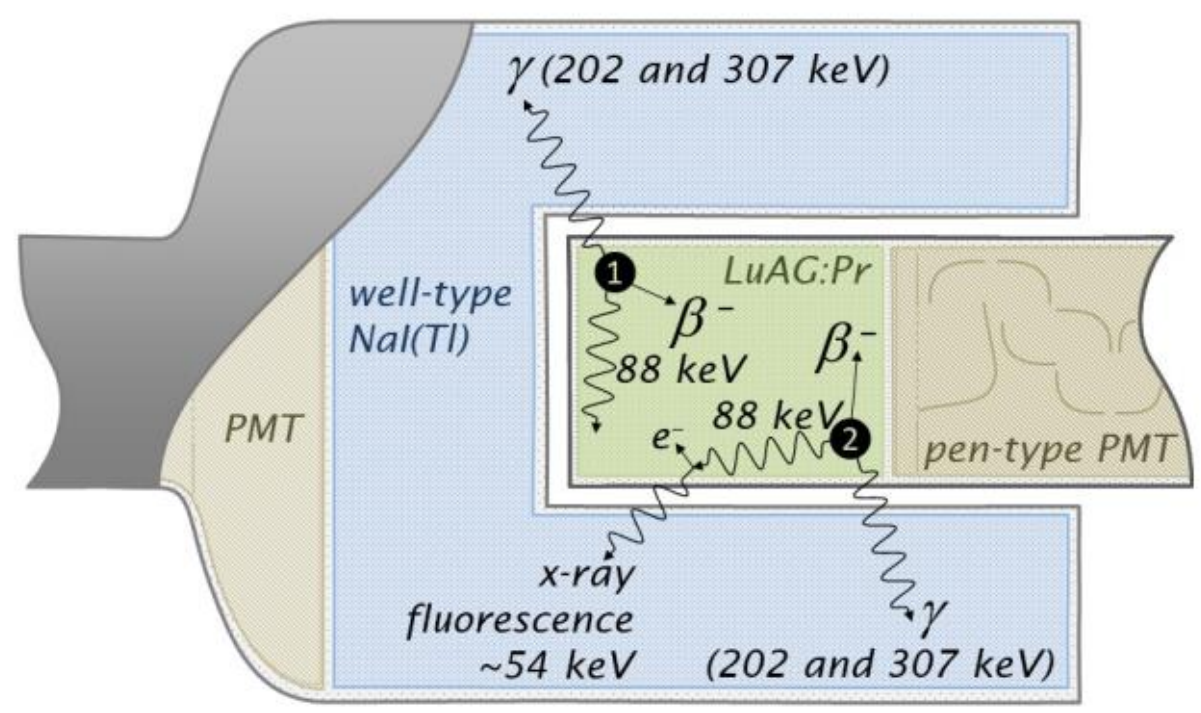

Figure 3. A longitudinal section view of the experimental set-up (not to scale). A LuAG:Pr scintillator coupled with a pen-type photomultiplier is fitted inside the well-type NaI(Tl) spectrometer. Two events of $\beta$ detection by the LuAG:Pr in coincidence with the NaI(Tl) are schematically reported: 1) 202 + 307 keV are detected by the NaI(Tl) and the full 88 keV + $\beta$ is detected by the LuAG:Pr; 2) again, 202 + 307 keV are detected by the NaI(Tl), the 88 keV generates a fluorescence x-ray ( 54 keV) which escapes the LuAG:Pr hence 88 – 54 = 34 keV + $\beta$ is detected by the LuAG:Pr.

Effects of the above mentioned nPR are also observable in Fig. 2 (Top). In fact, the spectra are linearly calibrated from channel to energy by taking into account the nPR characteristic of each scintillator as reported in [49, 51, 52]. For instance, the nPR of LSO:Ce at 60 keV is of the order of 83%, hence we calibrate the $^{241}$Am main peak at 59.54 × 0.83 = 49.4 keV. For comparison, the nPR of LuAG:Pr at 60 keV is 97%. The advantage of such non standard calibration is to preserve the actual response of the scintillator that can then be taken into account at a later stage in the construction of the response function. Moreover, the observed good linearity of such a calibration confirms that the data on nPR correctly apply to the specific scintillator sample in use. As a consequence of this calibration one can observe that the middle of the left edge of peak (a) is found approximately at 76 keV for LSO:Ce and at 87 keV for LuAG:Pr, both in agreement with their specific nPRs (86% and 98% respectively). For further information specifically concerning the escape probabilities of the various emissions in the case of LSO:Ce and LYSO:Ce, to some extent applicable as well to nearly equally dense LuAG:Pr, we refer to the dedicated work of Alva-S´anchez *et al.* [53].

A detailed analysis of the above self-scintillation spectra could lead to some capability of extracting the true $\beta$-spectrum of the dominant $\beta$-transition of $^{176}$Lu. However, the presence of the minor $\beta$-transition distorts to a certain degree the shape of the $\beta$-spectrum below 200 keV. Therefore one would need to make some assumptions on both $\beta$-shapes in order to disentangle each contribution from the other, discouraging one to proceed in this direction. Nonetheless, we could further analyze the self-scintillation spectrum of LuAG:Pr in search of electron capture related events as reported in Section IID.

The bottom part of Fig. 2 reports the emission of a 1 cm$^3$ LuAG:Pr scintillator as detected by a well-type 3$^{00}$×6$^{00}$ NaI(Tl) spectrometer (4$\pi$ geometry) and a 2$^{00}$×2$^{00}$ $CeBr_3$ spectrometer (<2$\pi$ geometry). Again, the measurements were performed inside a lead-castle and the residual environmental activity is also reported as the background spectrum for both NaI(Tl) and $CeBr_3$ spectrometers. From the spectra, it can be noted that the 88 keV $\gamma$-ray has a low probability to escape from the scintillator, not only because it interacts within the scintillators rather than escaping, but also because of its original low emission intensity of 14.5%. Alva-S´anchez *et al.* evaluated its escape probability from a 1 cm$^3$ LYSO:Ce scintillator as 5.1% [53]. Therefore, triggering on the escape of the whole de-excitation cascade of 88+202+307 keV in order to observe solely the corresponding $\beta$ in the scintillator will lead to a severe loss of counting efficiency that cannot be compensated for by simply increasing the size of the scintillator. Moreover, the limited energy resolution of the NaI(Tl) (8.5% FWHM at 662 keV) makes the 88+202+307 keV peak poorly resolved from those of 202+307 keV and 54+202+307 keV. This is not the case for the $CeBr_3$ (4% FWHM at 662 keV), which can resolve all three peaks. Nevertheless, by comparing the counting efficiency of the two spectra with NaI(Tl) and $CeBr_3$, the advantage of a 4$\pi$ geometry became evident. For the 202+307 keV peak, the counting efficiency increases at least 8 fold using a 4$\pi$ instead of a <2$\pi$ geometry.

Another aspect to be considered is that, with typical densities around 7 g/cm$^3$, Lu-containing scintillators strongly attenuate $\gamma$-rays. Therefore, according to NIST XCOM [54], a limit of about 1 cm$^3$ in the dimensions of the Lu-containing scintillator exists to allow about 20% of the 307+202 keV emissions to escape the scintillator.

Combining all of the above observations, the setup for the $\beta$-spectroscopy measurements has been designed to use the well-type NaI(Tl) spectrometer with a Lucontaining scintillator inside its well, coupled to a pentype PMT as schematically shown in Fig. 3. Every time that the NaI(Tl) detects 307+202 keV, the Lu-containing scintillator detects the $\beta$ in coincidence with the 88 keV emission. As a consequence, the $\beta$ detection is offset by and convolved with the 88 keV emission. This makes the use of a Lu-containing scintillator with satisfactory response in terms of energy resolution and proportionality crucial in order to reduce as much as possible the corresponding spectral smear. Unfortunately, the size of the NaI(Tl) well of 2/3$^{00}$ (1.69 cm) diameter implies the use of a 13 mm pen-type PMT, which, compared to larger PMT, typically offers modest quantum efficiency with significant limits in energy resolution.

In the end, we selected LuAG:Pr scintillators for the spectroscopic measurements of $^{176}$Lu $\beta$-decay. It has density of 6.7 g/cm$^3$ and an achievable energy resolution $\sigma$ < 14 keV at 662 keV (<5% FWHM). Its nPR at 10 keV is about 93%, and its specific activity is 215 Bq/cm$^3$. In comparison, LSO:Ce has a density of 7.4 g/cm$^3$, an achievable energy resolution of $\sigma$ = 22 keV at 662 keV (8% FWHM), nPR of 65% at 10 keV, and 295 Bq/cm$^3$ of specific activity. Custom sizes of LuAG:Pr were procured from Kinheng [55] and a demountable pen-type PMT assembly from Scionix [56]. The energy resolution measured with LuAG:Pr and the pen-type PMT was eventually $\sigma$ = 24 keV at 662 keV, consistent with the limited quantum efficiency of the pen-type PMT of about 20%. In fact, with a larger 2-inch PMT with quantum efficiency of 35%, we could measure $\sigma$ = 12 at 662 keV.

With the setup represented in Fig. 3, we acquired the self-scintillation with an 8×8×8 mm$^3$ LuAG:Pr scintillator in coincidence with the well-type NaI(Tl) detecting 202+307 keV. The measurement lasted for nearly 10 days and, with a counting rate of about 10 cps, $8.5 \times 10^6$ $\beta$decays were recorded. The stability of the signal against temperature and gain drifts was measured and showed limited fluctuations of less than 0.15%. The raw $^{176}$Lu $\beta$-spectrum is shown in Fig. 4. As mentioned earlier, this spectrum is not purely a $\beta$-spectrum but the convolution of the $\beta$ with the $^{176}$Hf de-excitation emission of 88 keV. This latter is detected by LuAG:Pr with an energy resolution of 24% FWHM, equivalent to $\sigma$ = 9 keV.

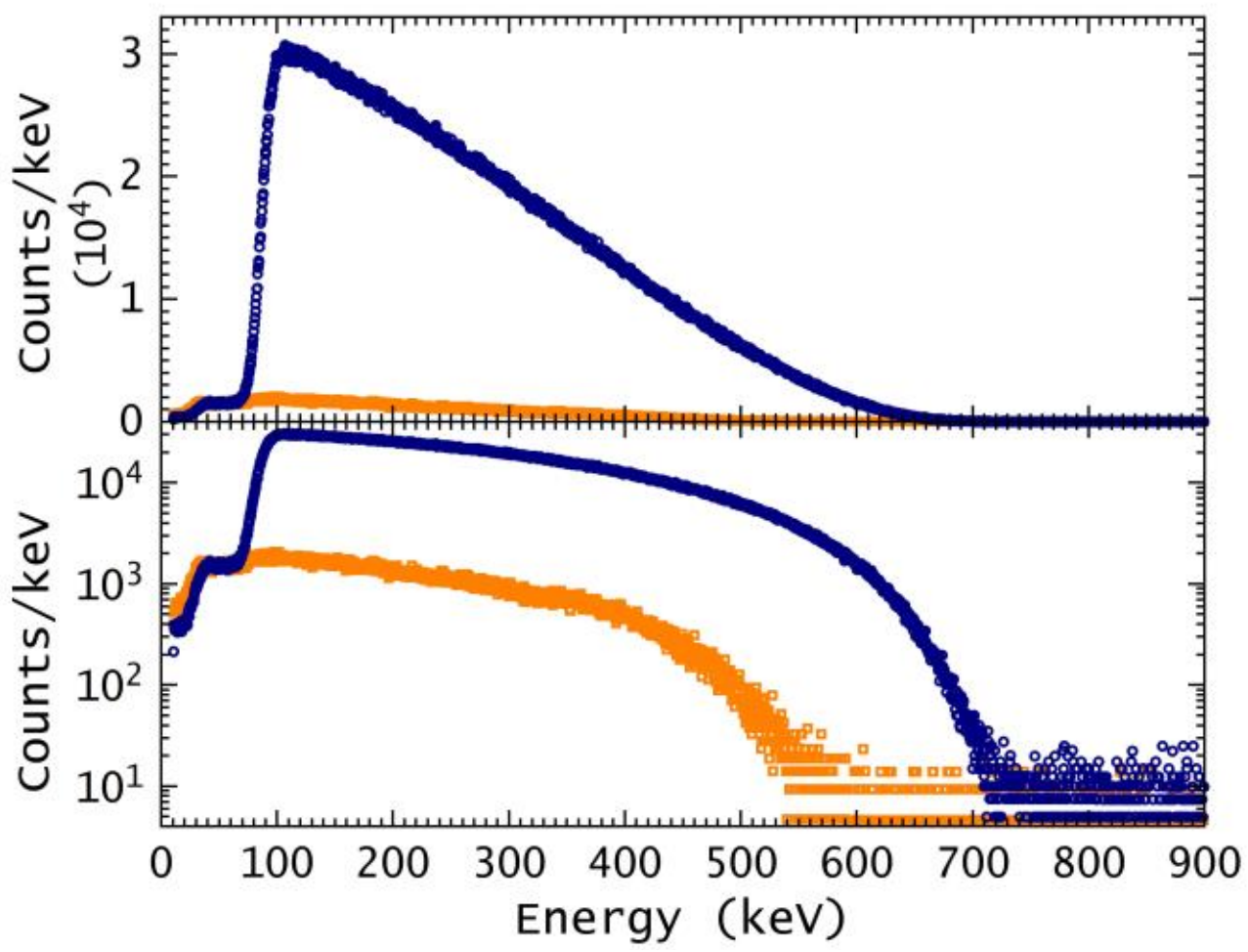

Figure 4. The raw spectrum of the dominant $^{176}$Lu $\beta$transition (blue circles) and the measured background (orange sqares), see text for details.

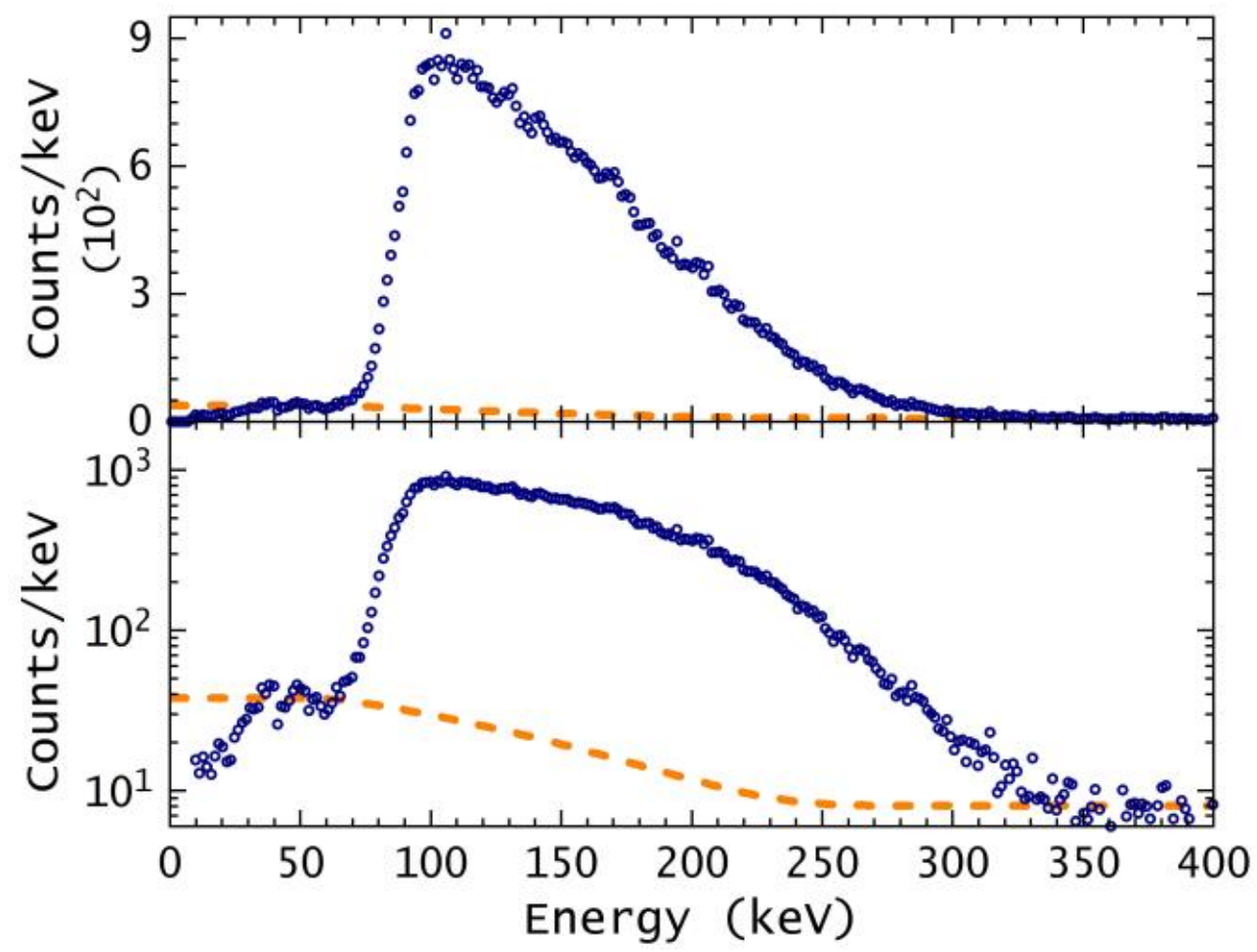

Figure 5. The raw spectrum of the minor $^{176}$Lu $\beta$-transition (blue circles) and the synthetically simulated background (orange dashed line), see text for details.

As in the self-scintillation spectra of Fig. 2, one can again observe the small sawtooth-shaped/step-like peak at 34 keV due to x-ray emissions. As schematically reported in the setup of Fig. 3, the 88 keV $\gamma$-ray can, via the photo-electric effect, generate a fluorescence x-ray which in turn can escape the scintillator. Alternatively but to an equivalent effect, the 88 keV (i.e. Hf($2^+$)) can undertake an internal conversion with x-ray emissions escaping the scintillator. Even with the limited energy resolution of the NaI(Tl) (see the overlap of the 202+307 keV and 54+202+307 keV peak in Fig. 2) one could set the gate to trigger out these unwanted events. However, they cannot be totally suppressed. In fact the 54 keV x-ray emission can also be absorbed in the dead layers surrounding the detectors, mostly aluminum, with equivalent consequences. Therefore, we instead triggered out almost entirely the 202+307 keV by setting the gate on the 54+202+307 keV and measured a spectrum dominated by x-ray emission + $\beta$ which is shown in Fig. 4. This spectrum can be subtracted from the 88 keV + $\beta$ spectrum in order to remove the unwanted events.

Besides the above, we also measured the second, least probable, $\beta$-transition of $^{176}$Lu and the raw data are reported in Fig. 5. This measurement was analogous to that of the first $\beta$-transition but with the gate set to trigger the 202+307+401 keV cascade. Here the count rate achieved was only 1.7 counts per minute and the acquisition had to last over one month to collect 83000 counts. A more severe gain drift with a drop approaching 1% was observed in the last week of the acquisition, quite possibly because of a change in the laboratory temperature, and we have corrected for it. Also, because of the low counting efficiency, this time the background due to the x-ray emission related events was not measured but rather simulated based on the expected shape of the $\beta$-transition. This simulated background is also reported in Fig. 5.

Another important measurement was performed with the setup of Fig. 3 with two pairs of LuAG:Pr scintillators: 8×8×2 mm$^3$ and 8×8×8 mm$^3$ cubic samples, and $\oslash 10\times 2$ mm$^3$ and $\oslash 10\times 10$ mm$^3$ cylindrical samples, with the aim to observe small detector effects [57]. These effects occur when the $\beta$ eventually escapes the scintillator and only part of its energy is detected causing a migration of counts from higher energy to lower energy in the shape of the $\beta$-spectrum. Despite the fact that the smaller samples have thicknesses about just twice the range of a 600 keV $\beta$ in LuAG:Pr, as shown in Fig. 6, each pair of spectra nicely overlap along the entire energy range. This result is further discussed in the following Sec. IIC along with the implementation of the experimental response function.

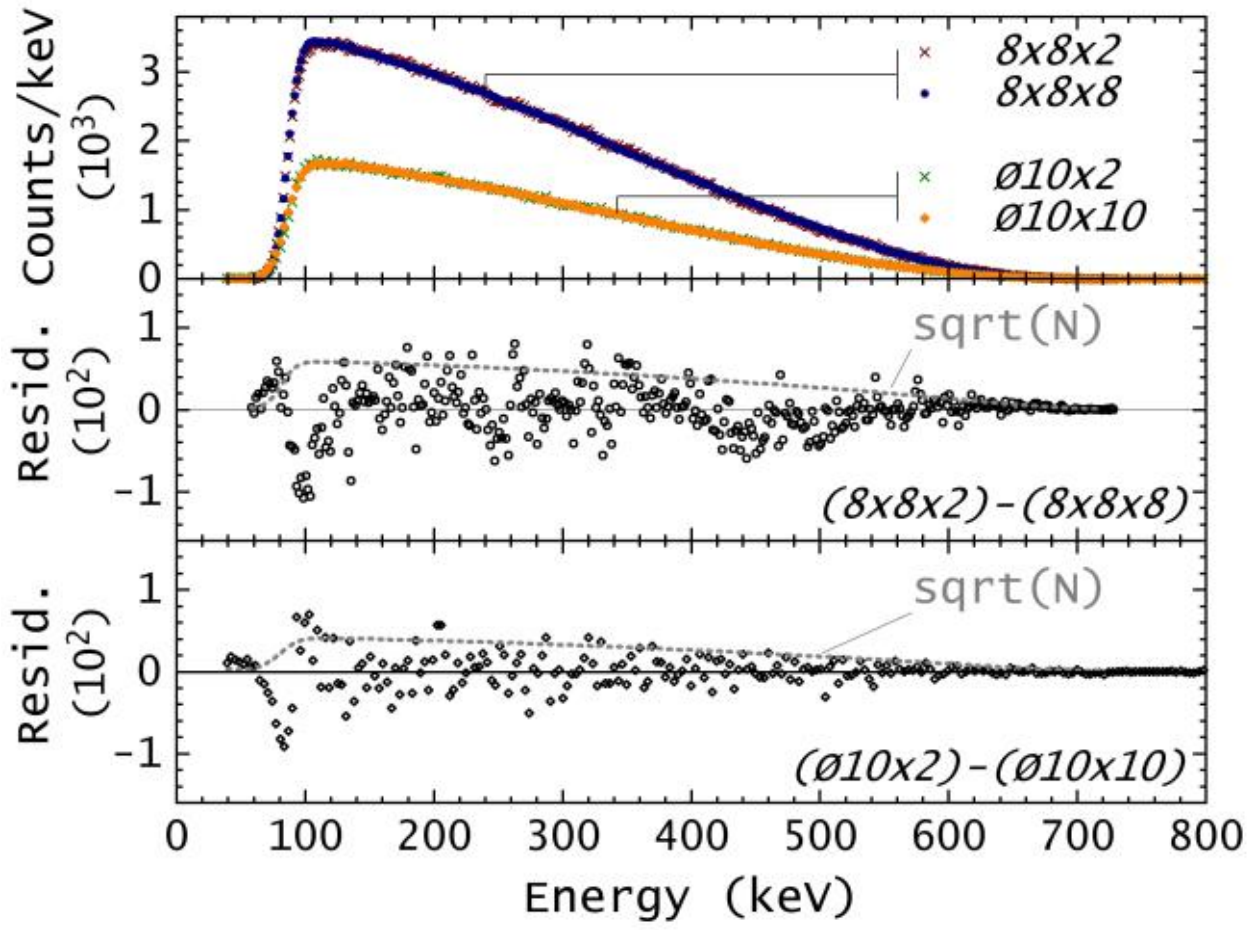

Figure 6. Comparison of the $\beta$+88 keV spectrum detected with two pairs of samples of LuAG:Pr of different dimensions, one pair corresponds to 8×8×2 mm$^3$ and 8×8×8 mm$^3$ and the other pair corresponds to $\oslash 10\times 2$ mm$^3$ and $\oslash 10\times 10$mm$^3$. Top: the normalised spectra. Center and bottom: the residuals for each pair of samples.

### C. $^{176}$Lu $\beta$-spectrum measurement results

In order to analyze the measured spectra we proceed by implementing the response function applicable to our experiment. We identified three sources of spectral smearing, namely: i) finite energy resolution, ii) nonproportionality of the response (nPR) and, iii), small detector effects. In addition, the $\beta$-particles have been measured in coincidence with 88 keV de-excitation emission. Therefore, the two signatures must be disentangled as well.

As explained above there were constraints on the maximum volume of the scintillator, hence we were particularly concerned by small detector effects. To evaluate them we used a fully empirical approach that eventually led us to consider negligible that source of distortion in the present experiment.

Small detector effects can be expected for $\beta$-particles emitted in close proximity to the surface of the scintillator that, as such, are more likely to escape the detector volume and hence avoid full energy deposition inside the scintillator. Moreover, during the process of slowing down to rest, $\beta$-particles can also emit bremsstrahlung radiation that, again, might escape the scintillator volume and hence detection. Since the higher the $\beta$ energy, the higher are both escape probabilities mentioned above, it can be expected that small detector effects appear in a $\beta$-spectrum as a migration of counts from the higher and middle energy channels to the lower energy channels.

According to NIST ESTAR [58], for LuAG:Pr one expects, in the continuous slowing down approximation (CSDA), a range of about 0.5 mm for 600 keV $\beta$-particles. Although this is just an approximation of the path actually travelled by an electron, advanced simulations generally appear in good agreement with it, as in the case of Prange *et al.* [59]. Now, the thickness of the smaller LuAG:Pr scintillators is 2 mm and therefore, considering the 0.5 mm electron range, about half of their volume is potentially affected by small detector effects. However, not even a minor migration of counts could be observed by comparing its $\beta$-spectrum to the one measured with the larger 8-mm-thickness-scintillators, as shown in Fig. 6.

The initial idea was to measure several scintillators with increasing thickness in order to quantify small detector effects vs the size of the scintillators, extrapolate the response of an ideally unaffected scintillator and then evaluate the correction for a real one. Substantial small detector effects were not expected based on previous experience with $^{138}$La as well as with experience in characterization of small scintillators in the frame of materials research (e.g. [48]), however the lack of any distinguishable effects was not anticipated for a dedicated experiment.

Such a lack of effects is rational for what concerns the bremsstrahlung. In fact, below 1 MeV, the energy loss of $\beta$-particles is dominated by collision processes rather than by bremsstrahlung emission and, according to NIST XCOM [54], one expects the radiation yield of a 600 keV $\beta$ to be of the order of 1% with an energy of about 17 keV. In turn, an x-ray of 17 keV has 99% absorption probability within 0.1 mm of LuAG:Pr and hence its overall escape probability becomes marginal.

Concerning the $\beta$ escapes, we could not find in the literature any equivalent experiment and so we could not compare our results with that of others. We consider all the above a quite interesting topic for further experimental and simulation-based investigations. On the one hand, it is rational to expect that a given scintillator material presents a sort of critical minimum volume below which that scintillator no longer absorbs all $\beta$-particles and their energy, hence causing distinguishable effects in the shape of the $\beta$-spectrum. On the other hand, we can conclude that that volume must be substantially smaller compared to the $\beta$ range and, therefore, in the context of the present experiment, no small detector effects need to be accounted for.

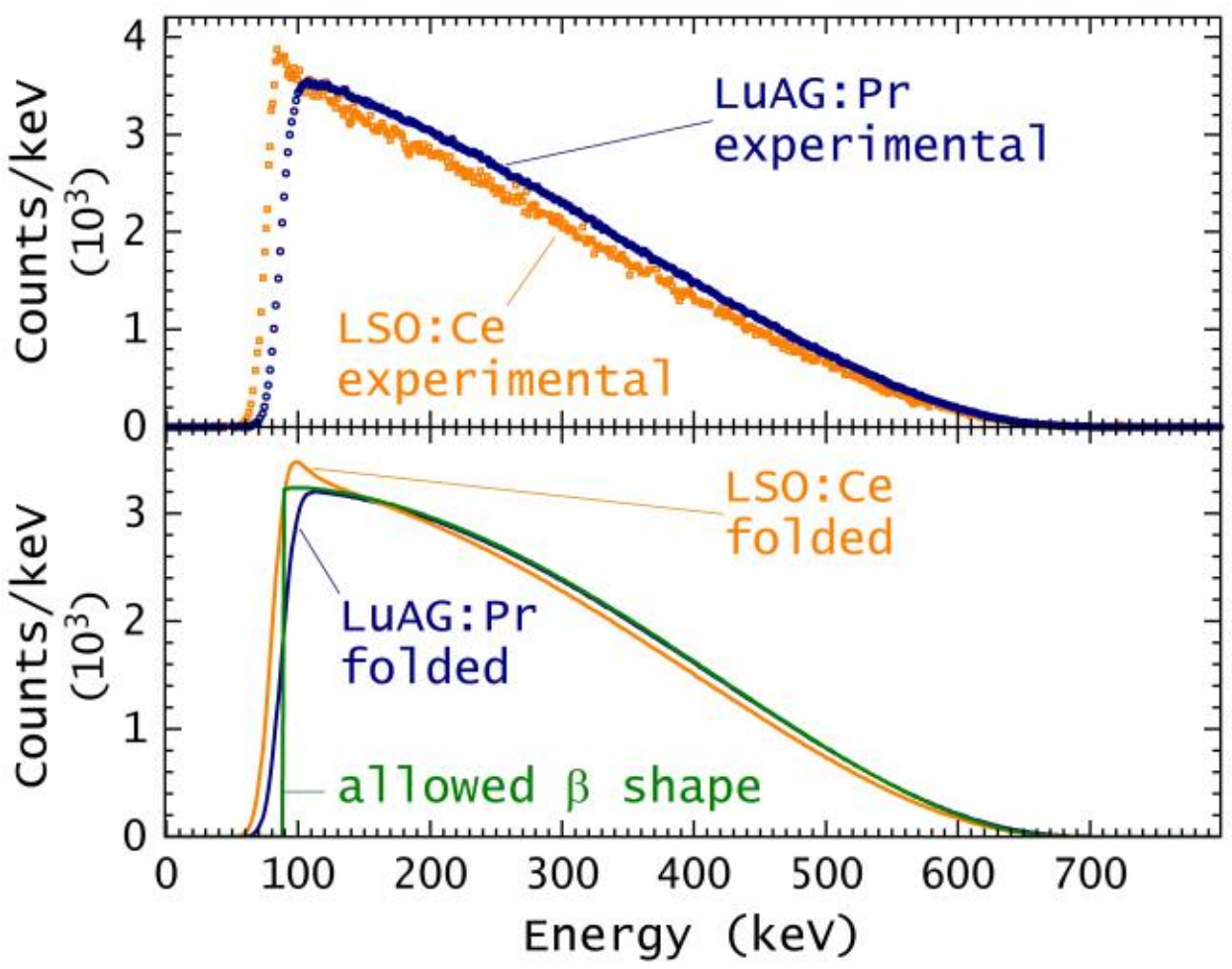

Figure 7. Top: Effects of nPR as seen in two spectra of $^{176}$Lu, one measured by nearly proportional LuAG:Pr and another measured by non-proportional LSO:Ce. Bottom: Implementation of the response function, which can be seen to reproduce the nPR effect of both scintillators.

The response function was implemented taking into account the finite energy resolution and the nPR. The energy resolution as a function of energy of the LuAG:Pr coupled with the low quantum efficiency pen-PMT was measured over a wide energy range with standard radioactive sources: e.g. $^{57}$Co, $^{60}$Co, $^{133}$Ba $^{137}$Cs and alike. By fitting the main x- and $\gamma$-ray peaks with Gaussian functions, we found that the energy resolution can be well represented in terms of $\sigma(E)$ = $0.93E^{0.50}$ with $E$ expressed in keV. Note that with a standard PMT, we measured $\sigma$ = $0.43E^{0.51}$, leaving room for further improvements. For each $\beta$-decay, both the corresponding $\beta$ and 88 keV de-excitation emission are coincidentally measured. However they follow distinct, uncorrelated scintillation processes and hence their energy resolutions combine in quadrature. With a $\sigma$ of about 9 keV, the energy resolution of 88 keV emission dominates the lower energy part of the spectrum.

Data on nPR of LuAG:Pr are available from [49, 50] covering, respectively, the energy ranges from 6 keV to 450 keV and from 0.1 keV to 30 keV. The two datasets present some discrepancies and, for instance, at 12 keV the reported nPR is 92% for [49] and 96% for [50]. However, both datasets show that LuAG:Pr is nearly proportional down to 10 keV and it is only below that energy that a stronger nPR starts to appear. In order to deal with the discrepancies, we averaged the two datasets into a smooth s-shaped curve. As discussed further on, LuAG:Pr being nearly proportional, the uncertainties in nPR have only minor effects.

The effects of nPR can be observed in Fig. 7 (Top) where the $\beta$-spectrum collected with LuAG:Pr is compared to one collected with the LSO:Ce during the preparatory phase of this work and with a different setup than that in Fig. 3. Because of the pen-type PMT used for LuAG:Pr, the two scintillators operate with nearly equivalent energy resolutions. However, their spectra present three main differences all due to nPR which for LSO:Ce is much stronger and already significant at energies of about 100 keV [49] . First of all, the LSO:Ce spectrum starts at lower energies than that of LuAG:Pr because of its stronger nPR which makes the detection of the 88 keV emission occur at about 76 keV. Secondly, the spectrum of LSO:Ce presents a much sharper peak than that of LuAG:Pr again because of its stronger nPR and consequent larger accumulation of counts towards the lower $\beta$ energies. Thirdly, its overall shape never overlap with the one of LuAG:Pr, making the experimental spectrum detected by LSO:Ce incompatible with the one detected by LuAG:Pr unless one takes into account the effects of nPR.

The response function was implemented in the form of a discrete matrix which generates the probability distribution to observe a $\beta$ with true energy $E$ at energy $E'$ in the measured spectrum. We firstly applied the response function to forward fold the allowed $\beta$-spectrum of $^{176}$Lu (i.e. C(W)=1) and the results are shown in Fig. 7 (Bottom) for the response functions of both LuAG:Pr and LSO:Ce. It can be observed that, in the case of LuAG:Pr, above 110 keV in the experiment scale (approximately above 20 keV of the $\beta$-spectrum) the folded spectrum overlaps the unfolded one. The same does not occur for LSO:Ce for which one can observe again the three features mentioned above for its experimental $\beta$-spectrum, namely: the starting of the spectrum at a lower energy; the much sharper peak; and the progressive accumulation of counts towards the lower energies. Albeit qualitatively, the forward folding with the LSO:Ce response function nicely confirms that the three features are indeed related to its much stronger nPR.

For LuAG:Pr, the forward folding of the allowed $\beta$spectrum was also used to determine the effects of uncertainty in nPR. We determined them using alternative evaluations of nPR and then looking at the total counts in the first 10 keV of the $\beta$-spectrum, where LuAG:Pr nPR is stronger. Note that in the folded spectrum of Fig. 7 (Bottom), taking into account the energy resolution $\sigma$ = 9 keV, the first 10 keV correspond to the energy range from 80 keV to 110 keV. Taking as reference the total counts for an ideally 100% proportional LuAG:Pr, the nPR of LuAG:Pr increased the total counts by 8.9% in the first 10 keV. For comparison, for LSO:Ce the total counts increased by 26.0%.

In addition, we generated two extra s-shaped curves of nPR, one using only the data from [49] and the other only with the data from [50]. In this case, the total number of counts from 80 keV to 110 keV increased by 8.2% and 9.0% respectively. We can then assess that uncertainties in the nPR have an influence of less than 1.0% in the first 10 keV of the $\beta$-spectrum. A similar approach for the energy range from 110 keV to the end point showed a negligible effect of 0.05%.

The response function was then used to unfold the measured $\beta$-spectrum into an approximation of the true spectrum by an iterative procedure. As mentioned earlier, the first 20 keV of the measured $\beta$-spectrum (i.e. the region from 88 keV to 88+20 keV) is offset and spread by the 88 keV of the de-excitation emission, with its nPR and its energy resolution $\sigma$ = 9 keV . The unfolding of this region of the spectrum presented a notable aspect. In fact, we found that uncertainties in the actual nPR at 88 keV have noticeable effects in the shape of the first 20 keV of the measured $\beta$-spectrum. In other words, the true nPR at 88 keV plays a crucial role and needs careful re-evaluation during the unfolding procedure. Note that uncertainties in the nPR at 88 keV do not directly propagate to the end point since there, that energy is summed to the energy of the $\beta$, and the resulting nPR becomes negligible. We found that the measured spectrum can be reproduced by re-folding the unfolded one if, in reason of the nPR, the 88 keV corresponds to 87 keV, consistently with the available data. For the unfolding procedure we applied the algorithm by Wortman and Cramer [60] and the results are reported in Fig. 8. We tested as well the unfolding algorithm of Magain [61] and found that the end point evaluation, reported further below, is independent of the unfolding algorithm. While the two algorithms generate nearly identical results, the spectrum unfolded by Magain's algorithm presented a minor increase of 800 counts in the first 5 keV, i.e. from 180100 counts to 180900 counts, i.e. a 0.5% increase, and a 0.3% increase in the first 10 keV.

Considering the effects of uncertainties in the nPR and intentionally altering the unfolded spectrum to observe how well it can be forward folded to the experimental one, we concluded that the unfolded spectrum can be considered a robust approximation of the true one from 20 keV to the end point. The shape of the first 20 keV of the unfolded spectrum presents however up to 1% relative uncertainty, and up to 3% in the first 3 keV.

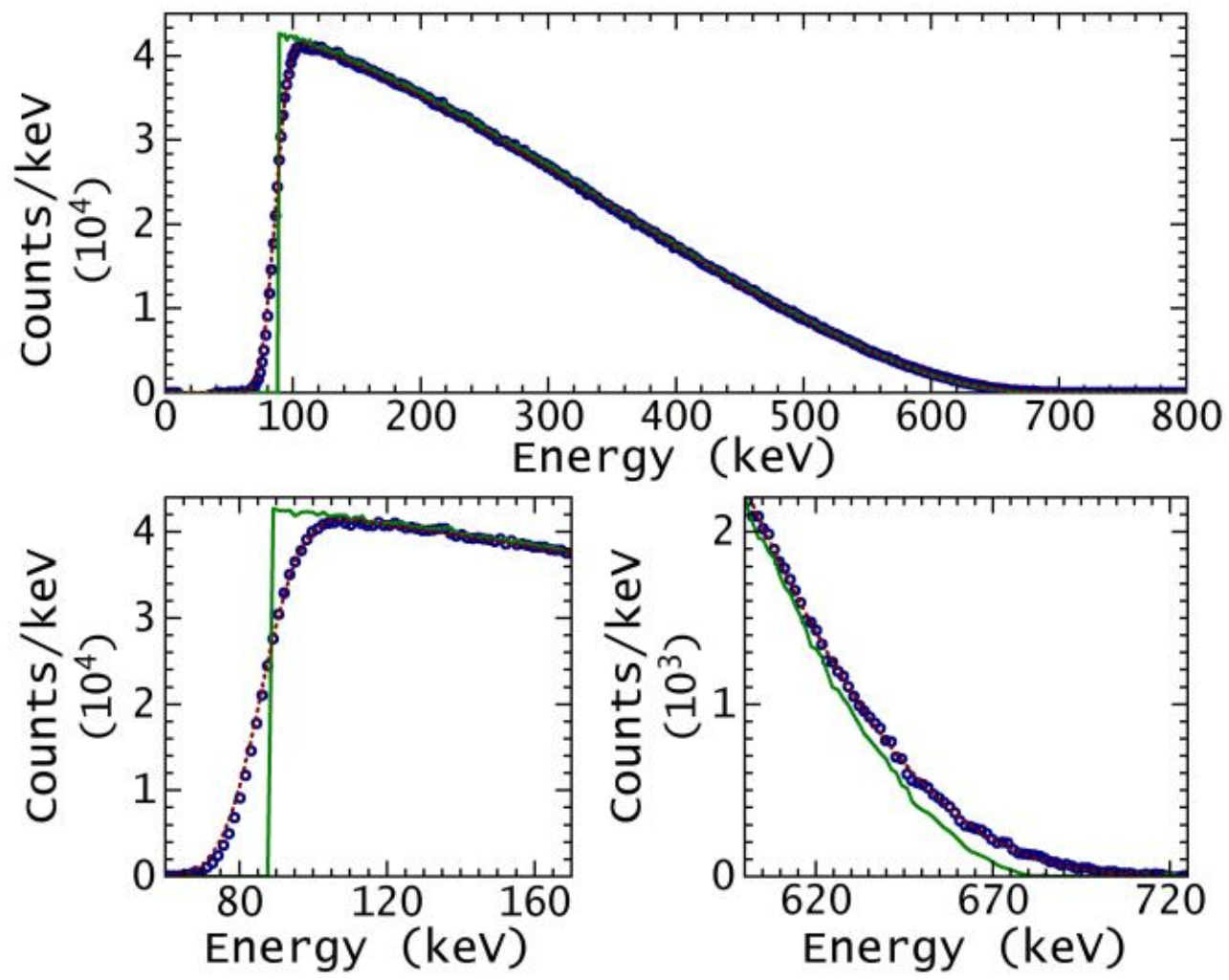

Figure 8. Top and bottom. The $^{176}$Lu main $\beta$-spectrum as measured by LuAG:Pr (blue circles), the same unfolded using the response function (green solid line) and the allowed $\beta$-spectrum, multiplied by the experimental shape factors, forward folded by the same response function (red dashed line). As seen in the two bottom plots, the spectral smearing, which in the case of LuAG:Pr is mostly due to its finite energy resolution rather than nPR, only occurs in the first 20 keV and toward the end point.

For an independent evaluation of the experimental end point of the main $\beta$-transition observed with LuAG:Pr, we lacked readily available shape factors and proceeded as follows. The unfolded experimental spectrum of Fig. 8, with its energy scale corrected for the offset produced by the 88 keV emission, was used together with a set of nine allowed $\beta$-spectra whose end points were increased from 591.0 keV up to 603.0 keV in steps of 1.5 keV. These spectra were obtained using BetaShape software [62, 63] and the end point values were chosen to cover a 12 keV range around the end point value of 597.0 keV derived from [25]. For each spectrum in the set of nine, we evaluated the corresponding experimental shape factors in the form

$$C(W)_{exp} = \frac{(dN/dW)_{exp}}{(dN/dW)_{comp}} = c\left(1 + aW + b/W + dW^2 + e/W^2\right) \quad (1)$$

where $C(W)_{exp}$ denotes the experimental shape factors, $W$ the total energy of the $\beta$-particle including its rest mass, $W = 1 + E/m_e$, $(dN/dW)_{exp}$ the experimental spectrum, $(dN/dW)_{comp}$ the computational spectrum for the allowed transition and $c$, $a$, $b$, $d$ and $e$ the fitting parameters. The form of $C(W)_{exp}$ was chosen to be able to fit the experimental shape factors corresponding to the most extreme end points in the set.

In order to evaluate the experimental end point, we proceeded by evaluating nine Kurie plots and observing how well they are fitted by straight lines, this in terms of the residual sum of squared (RSS) divided by the degrees of freedom (DoF), the RSS being also the quantity minimised by the fitting routine. The results are reported in Table I. By plotting the RSS/DoF vs the nominal end point and fitting the data points with a parabola, a minimum is found and hence the end point evaluated at 596.1(9) keV, well in agreement with the end point value obtained from the experimental Q value measured by the Penning trap mass spectrometry (i.e. 596.2(6) keV, see section IIIB, Table VI). The same procedure applied to the spectrum unfolded by Magain's algorithm, instead of Wortman and Cramer, led to the same end point value. We also found that an end point evaluation based on the coefficient of determination, instead of the RSS, led again to the same result.

The 0.9 keV uncertainty has two contributions adding in quadrature. The largest contribution corresponds to the uncertainty in locating the actual minimum in the RSS/DoF vs the nominal end point, conservatively evaluated as the half width of the step between the Kurie plots, i.e. 0.75 keV. The second contribution of 0.5 keV corresponds to the uncertainty in the channel to energy calibration.

Table I. Results of the procedure to determine the experimental end point of the first $\beta$-transition of $^{176}$Lu from its spectrum as measured by LuAG:Pr; RSS, residual sum of squares; DoF, degrees of freedom.

| Nominal end point (keV) | Fit end point (keV) | DoF | RSS/DoF |
|---|---|---|---|
| 591.0 | 591.73(15) | 392 | 0.153 |
| 592.5 | 593.11(14) | 393 | 0.137 |
| 594.0 | 594.49(13) | 394 | 0.125 |
| 595.5 | 595.86(13) | 395 | 0.118 |
| 597.0 | 597.23(13) | 396 | 0.116 |
| 598.5 | 598.56(13) | 397 | 0.122 |
| 600.0 | 599.90(14) | 398 | 0.136 |
| 601.5 | 601.24(15) | 399 | 0.160 |
| 603.0 | 602.58(16) | 400 | 0.192 |

An equivalent unfolding and end point evaluation procedure was applied to the second $\beta$-transition. Here the set of nine allowed $\beta$-spectra ranged from 190.0 keV to 202.0 keV again in steps of 1.5 keV and the shape factors did not include the term $e/W^2$. The results are reported in Table II. Again by plotting the RSS/DoF vs the nominal end points and fitting the data points with a parabola, the end point is found at 195.7(2.5) keV, once more in agreement with the value obtained from the experimental Q value (195.3(6) keV, see section IIIB, Table VI). The uncertainty of 2.5 keV, larger compared to the first $\beta$-transition, is due to a larger spread of the RSS/DoF data points and consequent larger uncertainty in locating their actual minimum.

Once the consistencies of the end point values obtained by $\beta$-spectroscopy and from the Penning trap Q value measurement were established, the latter value was taken to determine simpler forms of the shape factors because of its smaller uncertainty. For the first $\beta$-transition this was done using a spectrum with bin width of 0.3 keV while for the second transition we used again a 1.5 keV bin width. The results are shown in Fig. 9 and Fig. 10. The shape factors $C(W)_{exp}$ of the dominant $\beta$-transition are fitted by the parabolic equation in the form of

$$C(W)_{exp\ 1,1} =$$

$$= 2.032(8)[1 - 0.615(5)W + 0.178(9)W^2] \quad (2)$$

This equation however deviated from the experimental shape factors at energies below 10 keV. In order to cover that range a two parameter rational term can be added to (2), and the obtained best fit becomes

$$C(W)_{exp\ 1,2} = 1.967(11)\left[1 - 0.592(4)\,W + + 0.163(2)\,W^2 + \left(\frac{4.02(98)\ 10^{-3}}{1.029(9)\ W - 1}\right)^2\right] \quad (3)$$

The best fit of the Kurie plot obtained by applying $C(W)_{exp\ 1,2}$ matches the expected end point by intercepting the energy axis at 596.21(8) keV.

The shape factors of the second $\beta$-transition are also fitted by a parabolic equation in the form of

$$C(W)_{exp\ 2} =$$

$$= 1.81(89)[1 - 1.95(9)W + 1.4(3)W^2] \quad (4)$$

As seen in Fig. 10, towards the end point the experimental shape factors are steeply growing and the equation does not follows that behaviour. However, given the rather modest counting statistics in the measured spectrum we desisted from further analysis. The best fit of the Kurie plot intercepts the energy axis at 196.6(5) keV.

Table II. Results of the procedure to determine the experimental end point of the second $\beta$-transition of $^{176}$Lu from its spectrum as measured by LuAG:Pr; RSS, residual sum of squares; DoF, degrees of freedom.

| Nominal end point (keV) | Fit end point (keV) | DoF | RSS/DoF |
|---|---|---|---|
| 190.0 | 191.0(5) | 118 | 0.198 |
| 191.5 | 192.4(5) | 119 | 0.188 |
| 193.0 | 194.0(5) | 120 | 0.186 |
| 194.5 | 195.5(5) | 121 | 0.182 |
| 196.0 | 197.1(5) | 122 | 0.183 |
| 197.5 | 198.7(5) | 123 | 0.188 |
| 199.0 | 200.3(5) | 124 | 0.189 |
| 200.5 | 201.8(5) | 125 | 0.193 |
| 202.0 | 203.4(6) | 126 | 0.196 |

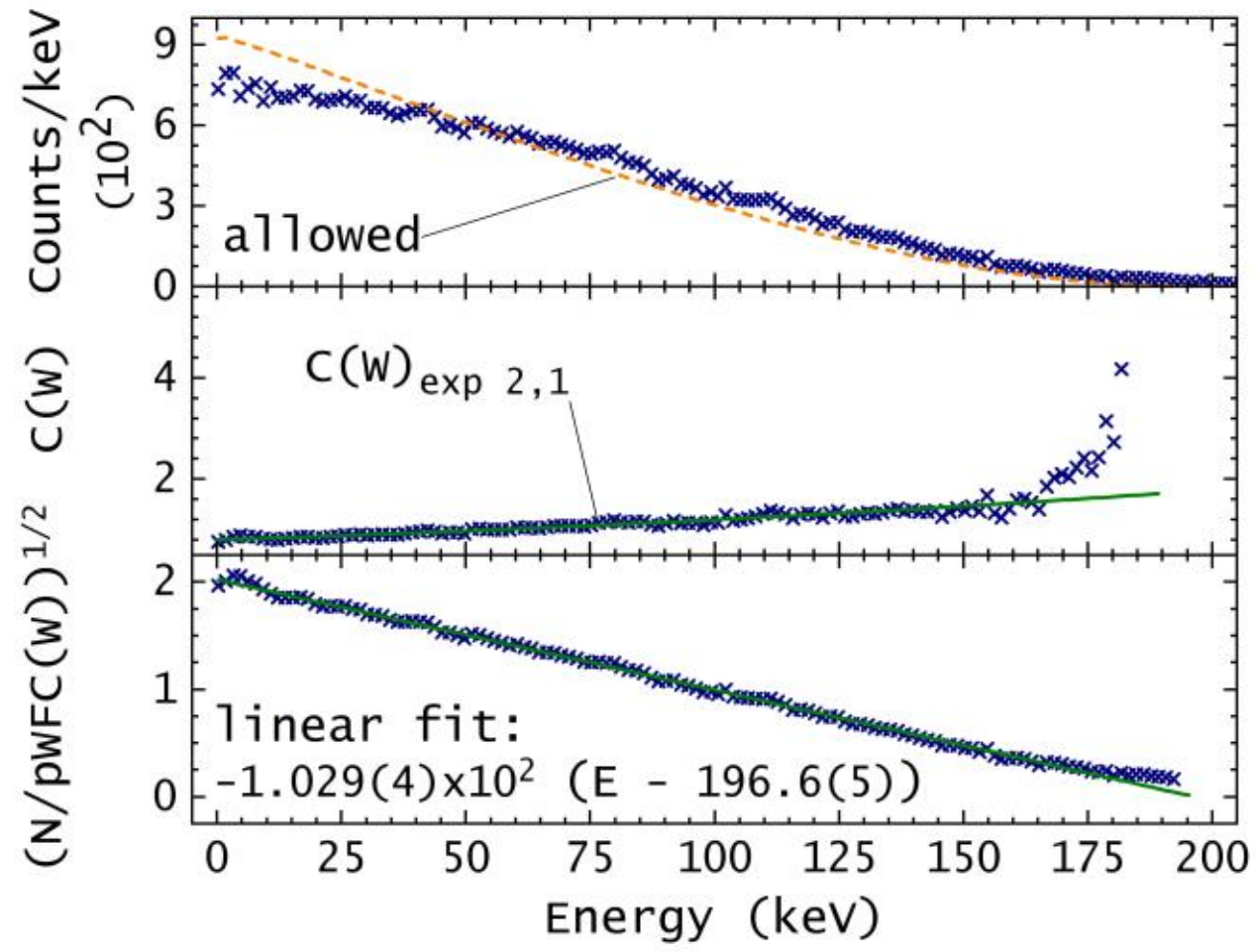

Figure 10. From top to bottom: the unfolded spectrum of the second $\beta$-transition of $^{176}$Lu, the corresponding experimental shape factors, and Kurie plot.

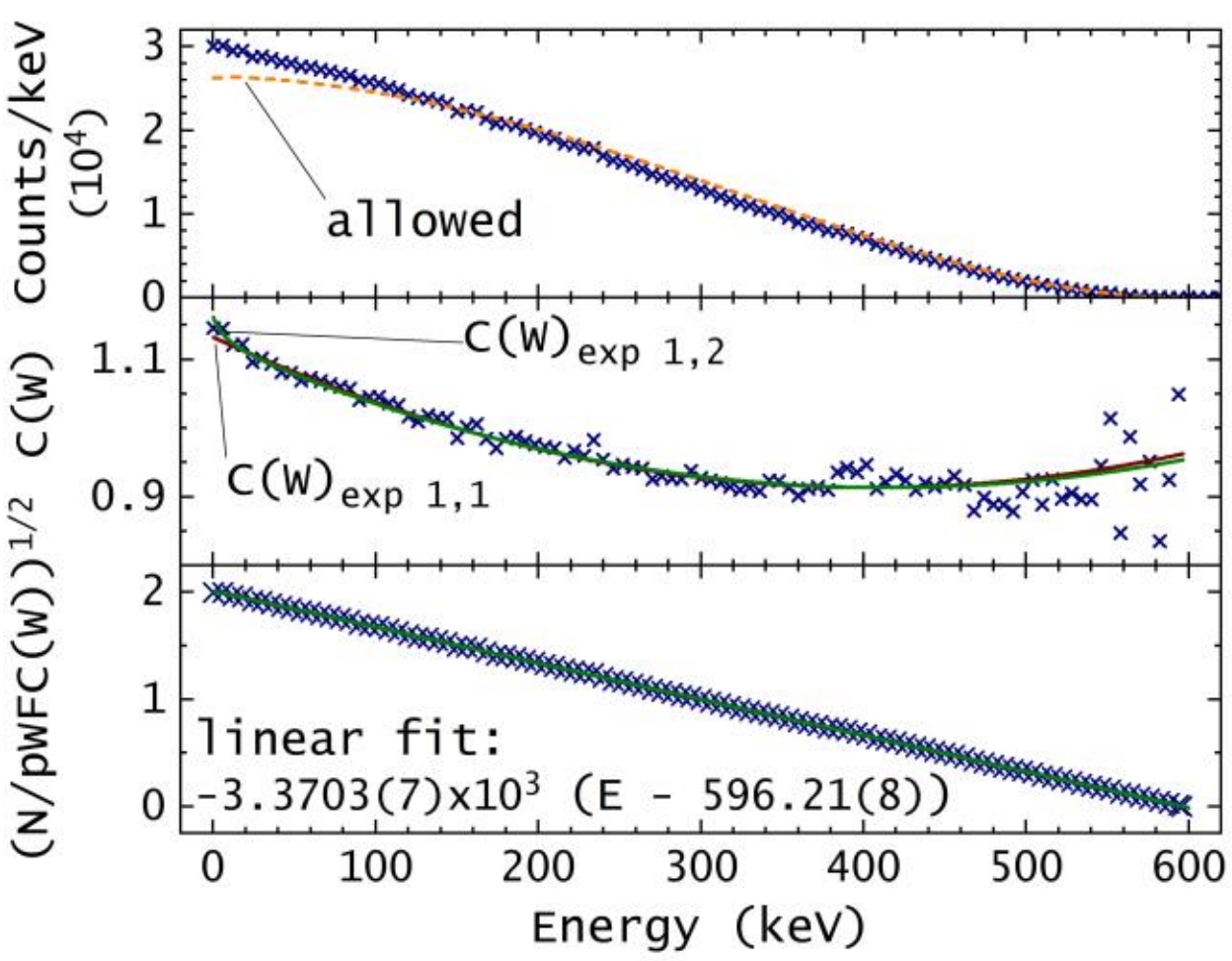

Figure 9. From top to bottom: the unfolded spectrum of the dominant $\beta$-transition of $^{176}$Lu, the corresponding experimental shape factors, and Kurie plot. Note that for better visualization, only 1 experimental data point (blue crosses) every 20 is plotted.

### D. Evaluation of $^{176}$Lu electron capture decay

As seen in Fig. 1, with a Q value of 108.9(8) keV (shortened to 109 keV in the text), there are two energetically allowed electron capture (EC) decays of $^{176}$Lu to $^{176}$Yb. One is a 5th forbidden non-unique transition to the $2^+$, 82.135(15) keV state of $^{176}$Yb (82 keV in the text). The second is a 7th forbidden non-unique transition to the $0^+$ ground state of $^{176}$Yb; both decays remain experimentally unobserved [28–30]. According to the limits by Norman *et al.* [28], the ratios of EC decay to $\beta$-decay are <0.45% and <0.36% for the 5th and 7th forbidden transition, respectively. Note that such limits compare to the <0.4% of the presently observed minor $\beta$-transition of $^{176}$Lu.

As observed in previous studies of $^{138}$La [33, 34], EC decays in self-scintillation spectra appear as peaks centered at the binding energies of the captured electrons in the daughter nucleus and with areas proportional to the respective capture probabilities. The energy at which the peaks are observed can be offset by de-excitation emissions, as it is in the case of $^{176}$Lu to $^{176}$Yb($2^+$, 82 keV). Capture probabilities of $^{176}$Lu are reported in Table III. They have been determined with the LogFT code [64] and the BetaShape code [63, 65]. None of them include the nuclear structure information that is required for these forbidden non-unique transitions. LogFT treats them as allowed while BetaShape treats them as forbidden unique transitions with the same change in total angular momentum (i.e. as 4th and 6th forbidden unique, respectively), which mainly explains the different capture probabilities.

As mentioned in Section IIB, the self-scintillation spectra of Fig. 2 (Top) would include counts originating from the EC decay branches of $^{176}$Lu. These spectra are in fact collected without specific coincidence conditions, hence all EC decays can accumulate there along with the $\beta$decays. On the other hand, the raw $\beta$-spectrum in Fig. 4 (as well as the one in Fig. 5) is virtually free from any EC decay event since the coincidence condition set for its acquisition, as described in Section IIB, allows collection of EC related counts only as random coincidences. In turn, these are so unlikely that they can be neglected. In fact, thanks to the large signal-to-noise ratio used to generate the gate (which also lasts just 6 $\mu$s), during that $8.3\times10^5$ s long acquisition only about 20 s were actually available for the random coincidence, so that even with an over-exaggerated EC branching of 10% the amount of possible random coincidences is less than five counts. In other words, the self-scintillation spectrum and the raw $\beta$-spectrum can be considered two extreme cases of, respectively, the presence of all possible EC counts and virtually no presence of EC counts.

Table III. Theoretical capture probabilities per electronic shell for $^{176}$Lu decay. The associated energy as detectable in the LuAG:Pr is also reported (in keV).

| | K | L | M | N | O | P |
|---|---|---|---|---|---|---|
| | $^{176}$Lu(7$^-$, gs) → $^{176}$Yb(2$^+$, 82.135(15) keV) | | | | | |
| LogFT | 0 | 0.508 | 0.492 | – | – | – |
| BetaShape | 0 | 0.0022 | 0.634 | 0.362 | 0.0013 | 3.3e-07 |
| Energy | 143.4 | 91.9 | 84.0 | 82.3 | 82.1 | 82.1 |
| | $^{176}$Lu(7$^-$, gs) → $^{176}$Yb(0$^+$, gs) | | | | | |
| LogFT | 0.532 | 0.347 | 0.1215 | – | – | – |
| BetaShape | 1.43e-06 | 0.3068 | 0.5256 | 0.1616 | 0.00593 | 1.86e-06 |
| Energy | 61.3 | 9.8 | 1.9 | 0.2 | 0.04 | 0.01 |

Fig. 11 (Top) reports the first 130 keV energy range of both the raw $\beta$-spectrum (same data as Fig. 4) and a self-scintillation spectrum which, below 130 keV, presents approximately the same number of counts of the $\beta$-spectrum, i.e. $1.45\times10^6$. No obvious shape difference can be observed between the two spectra. Moreover, by normalizing the areas of the two in order to take into account the different contribution of the escapes of x-ray emissions discussed in IIB, residual analysis shows no statistically significant differences between their shapes. It is only above 110 keV (incidentally just above the EC Q value) that the raw $\beta$-spectrum start to deviate from the self-scintillation one, as dictated by its sawtooth-shaped 88+$\beta$ peak discussed in IIB.

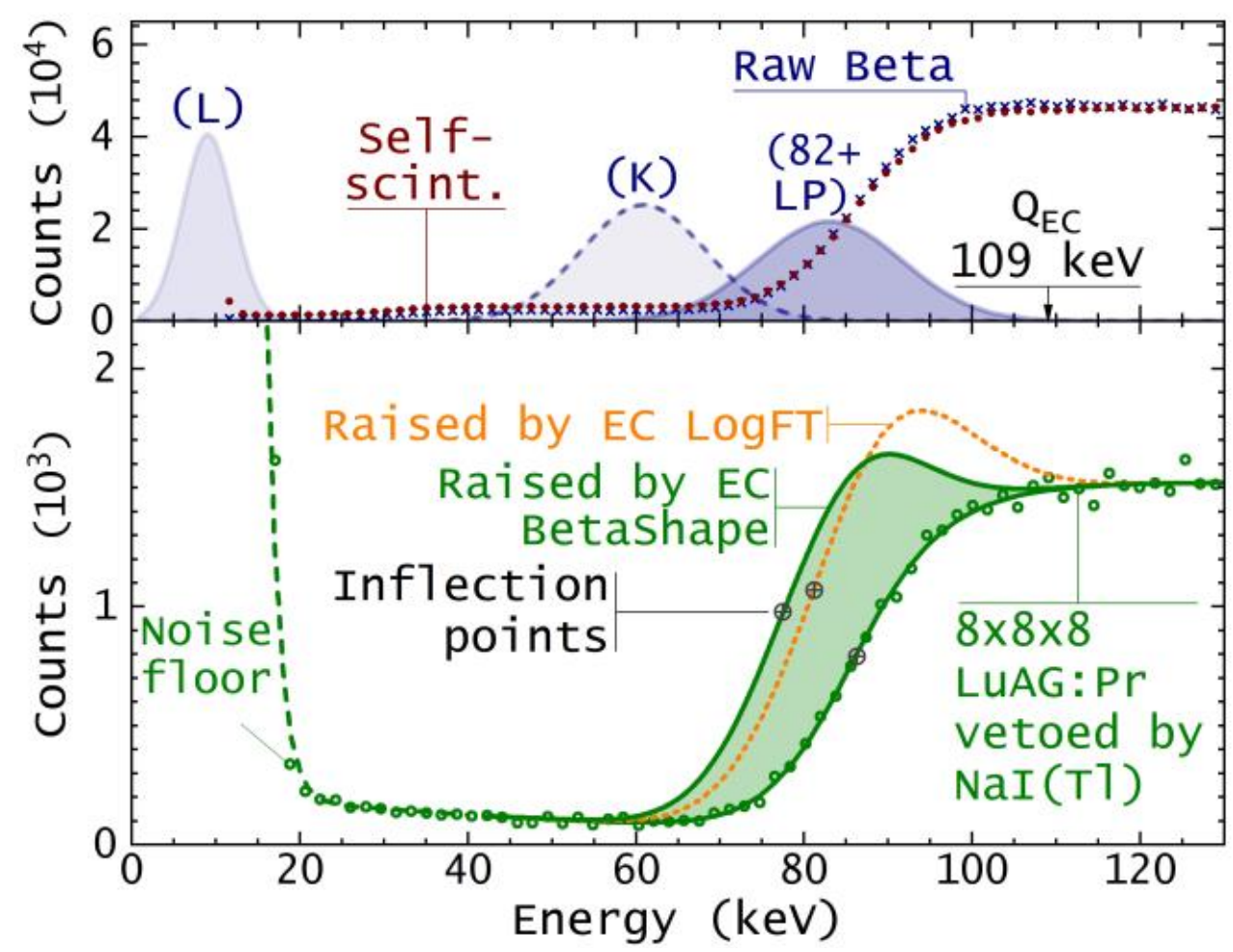

Figure 11. LuAG:Pr spectra below 130 keV used for the search of EC decays. Top. Raw $\beta$ and self-scintillation spectra (same spectra that in Fig. 2 and Fig. 4 respectively) and the expected EC peaks: (82+LP) represents all EC to the 82 keV level; (K) the K-shell captures to the ground state; (L) the L-shell captures to the ground state. Bottom. The self scintillation of LuAG:Pr collected in anti-coincidence with the NaI(Tl) and the effects corresponding to the detection of 10000 EC decays to the 82 keV level and for the capture probabilities of both BetaShape and LogFT.

Fig. 11 (Top) also reports the expected peaks for the two $^{176}$Lu EC decay branches as would be detected by the 8×8×8 mm$^3$ LuAG:Pr, i.e. taking into account its energy resolution and nPR. Note that because of the Q value of 109 keV and the 82 keV of the 2$^+$ level, K-shell captures are not considered since they violate energy conservation. The other energetically feasible electron captures of this branch, namely L-shell and higher shells up to the Pshell, have binding energies comparable or much smaller than the energy resolution of the LuAG:Pr with the pentype PMT ($\sigma$ = 8.5 keV at 82 keV) hence all EC to the excited level merge into the single, bell like, broad peak labelled (82+LP) in Fig. 11 (Top). The figure also shows that the present LuAG:Pr set up can observe EC to the ground state only through the K-shell peak, since the peaks of the L-shell and higher shell fall behind the 10 keV acquisition threshold.

Each of the three EC peaks in Fig. 11 (Top) is normalized to about 300000 counts, that roughly corresponds to the counts expected in the self-scintillation spectrum using the branching upper limits by Norman *et al.* [28].

For the K-shell peak, 300000 counts is a gross overestimation, being that the corresponding capture probability, according to BetaShape, is very small (see Table III). On the contrary, the peak of the EC decays to the 82 keV excited level is a convolution of all possible shell captures and only marginally affected by their actual probabilities. As seen, the 82+LP peak overlaps with the left-edge of the 88+$\beta$ peak. By fitting the energy range between 60 keV and 110 keV of both the self-scintillation and the raw $\beta$-spectra with a sigmoidal function, one can then look at the function inflection points to reveal the presence of EC to the 82 keV level. The value found for the inflection points are 86.28(5) keV and 86.16(4) keV for the self-scintillation and the raw $\beta$-spectra respectively. The presence of EC would have shifted the inflection point of the self scintillation to lower energy, the reverse of the above, hence the 0.12 keV observed difference cannot be considered significant for EC detection and rather arises from systematic uncertainties related to, e.g., energy calibration or gain stability (±0.15%). Observation of EC decays cannot be claimed.

To search further for EC decays, a dedicated acquisition in anti-coincidence was carried out, again with the set up of Fig. 3. This time LuAG:Pr counts were vetoed when coincident counts in the NaI(Tl) were occurring. At energies below 130 keV, such acquisition provided a $\beta$ self-scintillation spectrum dramatically reduced by nearly 70 times, hence substantially increasing the signal to noise ratio in favor of EC detection. Note that, on the whole energy range up to 1.2 MeV, the count rate in LuAG:Pr decreased only to about 1/3, consistent with its high detection efficiency. The LuAG:Pr spectrum vetoed by NaI(Tl) is reported in Fig. 11 (Bottom). Fitted again with a sigmoidal function, an inflection point of 86.2(2) keV was found, consistent with the fit values previously obtained, and characterised by a larger error because of its lower statistics, despite its acquisition lasting 0.61 Ms (170 hours) and collecting $19.2 \times 10^6$ counts. One can logically conclude that even in the vetoed spectrum, the amount of possible EC captures is so marginal that it cannot emerge above statistical noise.

Note that the lack of EC detection cannot be associated to a lack of detection efficiency for the 82 keV. According to Ott *et al.* [29], its photon emission probability is in fact 0.125, close to that of the 88 keV level of $^{176}$Hf (i.e. 0.149). Due to this small emission probability, adapting the work of Alva-S´anchez *et al.* [53] to the present 8×8×8 mm$^3$ LuAG:Pr, and reasonably assuming full efficiency for internal conversions, the detection efficiency of the 82 keV level is close to 100%. The actual evaluation led to 99.2%. In other words, considering that each EC decay to the 82 keV level is detected introduces only a marginal uncertainty.

Fig. 11 (Bottom) shows as well the expected effect of 10000 counts of EC decay to 82 keV level superimposed to the vetoed spectrum, this for both EC probabilities of BetaShape and LogFT (Table III). As seen, these 10000 counts would have obviously deformed the vetoed spectrum and moreover they represent about 30 times less the number of counts expected using the upper limit of Norman *et al.* [28]. In terms of inflection points, 10000 counts correspond to 81.2 keV and 77.5 keV for the LogFT and BetaShape capture probabilities, respectively. Based on the error in the fit parameters, a 0.5 keV shift of the inflection point would have been observed. In reverse, a 0.5 keV left-shift is associated with about 425 EC counts which, given the vetoed acquisition settings, would have occurred along with $67.3 \times 10^6$ $\beta$-decays. An upper limit for the EC decays to the 82 keV level of $^{176}$Yb can be obtained from the ratio:

$$\frac{EC\ decay}{\beta\ deacay} = \frac{425}{67.3 \times 10^6} = 6.3 \times 10^{-6}$$

Alternatively one can apply the counting statistics approach of Norman *et al.* [28] in the energy range from 60 keV to 110 keV where the vetoed spectrum consists of 22300 counts and the $1\sigma$ upper limit is then sqrt(2 × 22300) = 211 counts so that the above ratio EC decays over $\beta$-decays is then < $3.2 \times 10^{-6}$.

The energy range of interest to observe the K-shell captures to the ground state is from 40 keV to 80 keV, where 3300 counts are present in the vetoed spectrum. An analysis equivalent to the one above is however of little significance for the setting of an upper limit because of the substantially negligible K-shell capture probability. On the other hand, the upper limit found for the 5$^{th}$ forbidden EC to the 82 keV applies as well to the 7$^{th}$ EC to ground state, the latter being, in general terms, much less likely than the former.

## III. Q VALUE MEASUREMENT

### A. Penning trap apparatus and measurement

The $^{176}$Lu Q value and mass excess measurements were performed at the Low Energy Beam and Ion Trap (LEBIT) facility at the National Superconducting Cyclotron Laboratory (NSCL) [66]. LEBIT is a Penning trap mass spectrometry (PTMS) facility that was designed for precise mass measurements with short-live radioactive isotopes produced via projectile fragmentation by the coupled cyclotron facility at the NSCL. However, it also includes two offline ion sources—a Laser Ablation Source (LAS) [67] and a Thermal Ion Source (TIS)—that are used to produce ions of stable isotopes for calibration and reference purposes. These sources have also been used to produce long-lived isotopes for nuclear and neutrino physics studies. A schematic of the components of LEBIT relevant to the measurements described here is shown in Fig. 12.

In this work, the LAS was used to produce singly-charged $^{176}Lu^+$, $^{176}Hf^+$, and also $^{176}Yb^+$ ions from sheets of approximately 25 × 12 × 1 mm thick, naturally abundant lutetium, hafnium, and ytterbium samples. Sheets of two different materials were mounted side by side on the LAS target holder, which is connected to a stepper motor to enable selective production of ions of two different isotopes during a single experimental run.

After production in the LAS, ions are transported to a beam cooler/buncher [68] consisting of a helium gas filled RFQ ion guide and trap that is used to produce low emmittance, short duration ion bunches. The ion bunches are then transported to the Penning trap where they are captured and the measurement is performed.

The LEBIT Penning trap [69] has a hyperbolic trap structure, housed within a 9.4 T superconducting solenoidal magnet. LEBIT uses the Time of Flight-Ion Cyclotron Resonance (TOF-ICR) technique [70] to precisely measure the cyclotron frequency of ions within the trap:

$$f_c = \frac{qB}{2\pi m} \tag{5}$$

For this experiment, the Ramsey excitation scheme was used [71–73]. Ions within the trap are driven with two time-separated quadrupolar radiofrequency (rf) pulses with a frequency close to $f_c$. The ions are then released from the trap toward a microchannel plate detector (MCP), where their time-of-flight from the trap to the detector is measured. This measurement is repeated for a series of ion bunches, with the frequency of the rf pulse systematically varied around $f_c$. The resulting times-offlight produce a resonance curve, an example of which is shown in Fig. 13 for $^{176}Lu^+$. The central minimum corresponds to $f_c$ and can be obtained from a fit to the theoretical line shape.

To account for time-related frequency shifts, cyclotron frequency measurements, like the one presented in Fig. 13, are alternately taken for the two isotopes in the LAS. By measuring the frequency of ion one at time $t_1$, ion two at time $t_2$, and ion one again at time $t_3$, the two cyclotron frequency measurements for ion one, $f_{c1}(t_1)$ and $f_{c1}(t_3)$, can be linearly interpolated to find $f_{c1}(t_2)$ at time $t_2$. This is then used to find the cyclotron frequency ratio of the two ions:

$$R = \frac{f_{c1}(t_2)}{f_{c2}(t_2)} = \frac{m_2}{m_1}. \tag{6}$$

The alternating measurements are repeated a number of times (in this work up to 94 times—see Table IV) and the average ratio, $\bar{R}$, and Birge Ratio [74] for the data set are calculated. When the Birge Ratio is greater than one, the uncertainty in $\bar{R}$ is inflated by the Birge Ratio to account for possible underestimation of the statistical uncertainty. The average ratio, $\bar{R}$ can then be used to directly calculate the Q value of the decay, as discussed in Section IIIB.

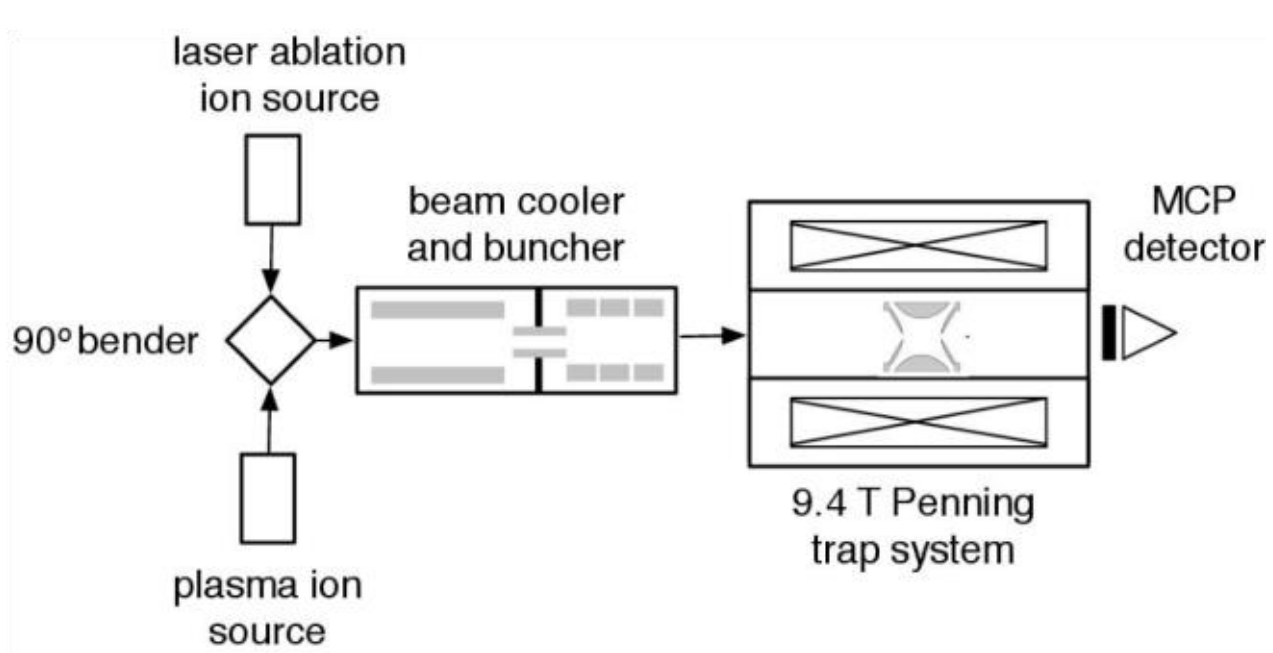

Figure 12. A schematic of the sections of the LEBIT beamline used in this work.

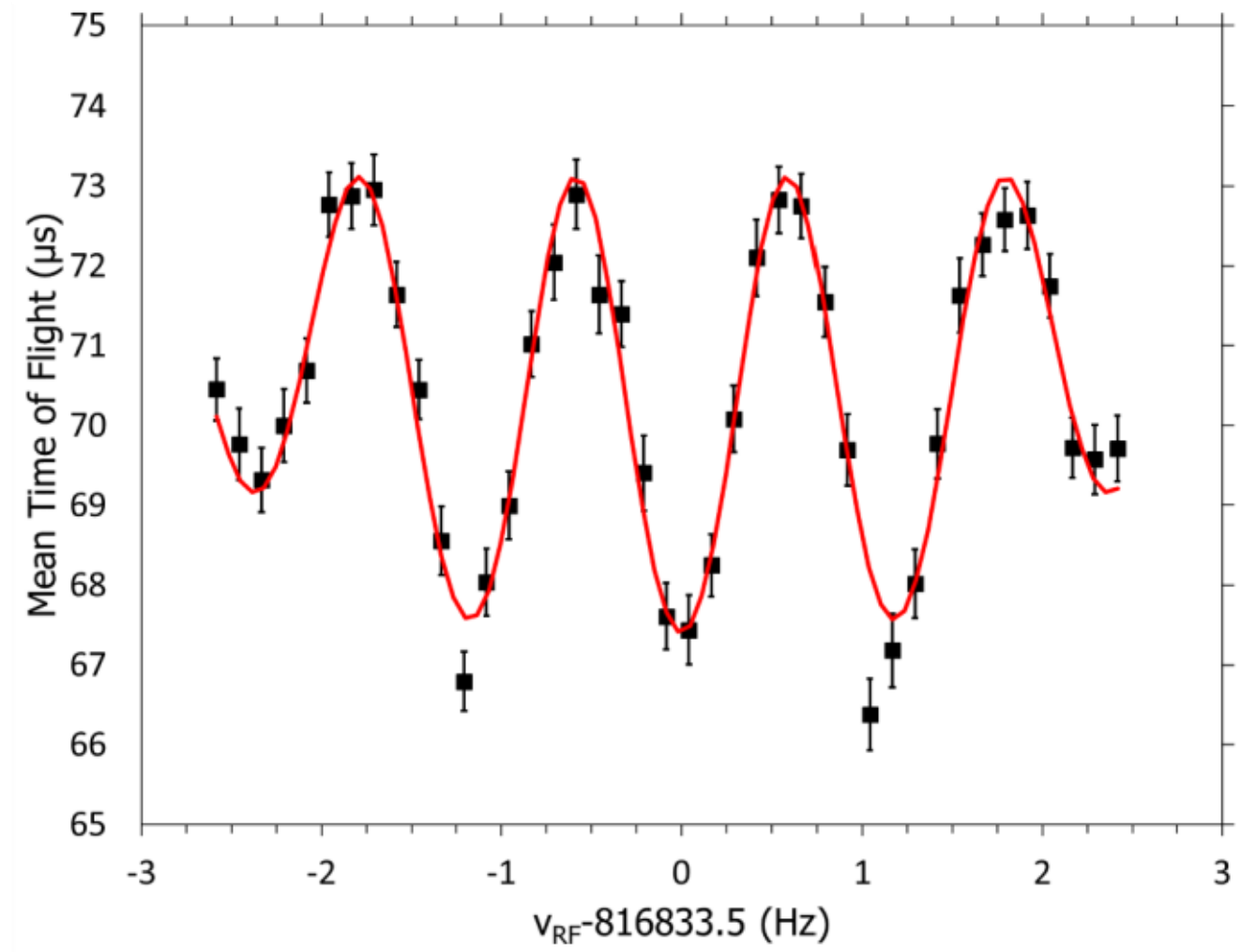

Figure 13. A 1.0 s Ramsey ion cyclotron resonance for $^{176}Lu^+$. The solid red line is the theoretical fit to the data.

Table IV. Measured cyclotron frequency ratios for combinations of $^{176}Lu^+$, $^{176}Hf^+$, and $^{176}Yb^+$ ions among themselves. N is the number of individual ratio measurements contributing to the average, $\bar{R}$. The uncertainties for $\bar{R}$, shown in parentheses, have been inflated by the Birge Ratio, BR, when
BR > 1.

| Ratio | Ion Pair | N | BR | $\bar{R}$ |
|---|---|---|---|---|
| (i) | $^{176}Lu^+/^{176}Hf^+$ | 94 | 1.1 | 0.999 992 725 1(41) |
| (ii) | $^{176}Lu^+/^{176}Yb^+$ | 62 | 1.1 | 0.999 999 335 5(47) |
| (iii) | $^{176}Yb^+/^{176}Hf^+$ | 38 | 1.2 | 0.999 993 380 4(26) |

### B. $^{176}$Lu $\beta$-decay Q value

The three average cyclotron frequency ratios measured in this work are given in Table IV. These ratios were used to directly obtain the Q value of the decay between relevant parent and daughter nuclides i.e. the $\beta$-decay of $^{176}$Lu to $^{176}$Hf, the EC-decay of $^{176}$Lu to $^{176}$Yb, and the 2$\beta$-decay of $^{176}$Yb to $^{176}$Hf, via

$$Q = (M_p - M_d)c^2 = (M_d - m_e)(1 - \bar{R})c^2, \tag{7}$$

where $M_p$ is the mass of the parent atom, $M_d$ is the mass of the daughter atom, $m_e$ is the mass of the electron, and $\bar{R}$, previously defined in Eqn. (6), is such that ion 1 refers to the parent, and ion 2 the daughter. The ionization energies of the parent and daughter atoms are two orders of magnitude smaller than the statistical uncertainty, and are therefore ignored in this work.

The main goal of this work was to precisely determine the $^{176}$Lu $\beta$-decay Q value. From the data in Table IV, this Q value can also be obtained by taking the product of ratios (ii) and (iii) to obtain an independent measurement of $\bar{R}$ for $^{176}$Lu$^{+}$/$^{176}$Hf$^{+}$ and again using Eqn. (7). All of the obtained Q values, along with the values from the AME 2020 [25], are given in Table V. The measured Q values are in good agreement with the AME 2020, but are more precise, with all of the new measurements having an uncertainty < 1 keV.

The $^{176}$Lu $\beta$-decay Q value can be used to determine the end point energies for the primary (99.61%), and secondary (0.39%) branches to the 596.82(5) keV $6^{+}$ and 997.73(6) keV $8^{+}$ levels, respectively in $^{176}$Hf [27]. The resulting end point energies are given in Table VI.

Table V. Q values for $^{176}$Lu $\beta$-decay or EC-decay and $^{176}$Yb 2EC-decay calculated from cyclotron frequency ratios listed in Table IV. The calculated Q value is listed along with the AME 2020 value [25] and the difference $\Delta Q = Q_{\text{LEBIT}} - Q_{\text{AME}}$.

| Decay | Ratio | Q value (keV) LEBIT | AME | $\Delta Q$ |
|---|---|---|---|---|
| | (i) | 1192.28(67) | | -1.8(11) |
| $^{176}$Lu($\beta^{-}$) | (ii)/(iii) | 1193.78(88) | 1194.09(87) | -0.3(12) |
| | Avg. | 1193.03(55) | | -1.1(10) |
| $^{176}$Lu(EC) | (ii) | 108.90(76) | 109.0(12) | -0.1(14) |
| $^{176}$Yb($2\beta^{-}$) | (iii) | 1084.88(43) | 1085.1(15) | -0.2(15) |

Table VI. $\beta$-spectrum end point energies from this work deduced with the data from [27] for the primary and secondary decays of $^{176}$Lu to $6^{+}$ and $8^{+}$ levels in $^{176}$Hf.

| Daughter level $J^{\pi}$ | $E^{*}$ (keV) | Branch Strength | $\beta$ end point (keV) |
|---|---|---|---|
| $6^{+}$ | 596.82(5) | 99.61 % | 596.21(55) |
| $8^{+}$ | 997.73(6) | 0.39 % | 195.30(55) |

Table VII. Mass excesses, ME, for $^{176}$Lu and $^{176}$Hf, obtained from the ratios listed in Table IV. The results are compared to those listed in the AME2020 [25]. The column ΔM is calculated as $\text{ME}_{\text{LEBIT}} - \text{ME}_{\text{AME}}$

| Nuclide | ME (keV/$c^2$) LEBIT | AME | ΔM (keV/$c^2$) |
|---|---|---|---|
| $^{176}$Lu | -53 382.42(76) | -53 382.3(12) | -0.1(14) |
| $^{176}$Hf | -54 576.20(43) | -54 576.4(15) | 0.2(15) |

### C. $^{176}$Lu and $^{176}$Hf atomic masses

Mainly from a PTMS measurement at Florida State University [75], the atomic mass of $^{176}$Yb is known to a precision of 14 eV/$c^2$ [25]. Thus, using ratios (ii) and (iii) in Table IV and the atomic mass of $^{176}$Yb, it was possible to determine the absolute masses of $^{176}$Lu and $^{176}$Hf, using the equation

$$M = (M_{\text{Yb}} - m_e)\bar{R}^{-1} + m_e, \qquad (8)$$

where $M$ is the mass of atom to be determined ($^{176}$Lu or $^{176}$Hf), $M_{\text{Yb}}$ is the mass of $^{176}$Yb, and $\bar{R}$ is the appropriate ratio, (ii) or (iii) from Table IV, for $^{176}$Lu or $^{176}$Hf, respectively. The resulting masses are reported in Table VII as mass excesses, which are calculated from

$$\text{ME} = (M - A) \times 931494.10242(28)(\text{keV}/c^2)/\text{u}, \qquad (9)$$

where $A$ = 176 is the mass number of the ion of interest and the conversion factor between keV/$c^2$ and u is from Ref. [76]. The mass excesses we obtain in this work are in good agreement with the AME2020 values, which come primarily from $(n,\gamma)$ and $\beta^{-}$-decay measurements that link $^{176}$Lu and $^{176}$Hf to $^{174}$Yb, which has been precisely measured with the Florida State University Penning trap [75]. The results are in good agreement with other, less precise, measurements performed with the TRIGA-TRAP Penning trap [26]. Here we have reduced the uncertainties by factors of 1.5 and 3 for $^{176}$Lu and $^{176}$Hf, respectively, and in each case the uncertainties are both now < 1 keV/$c^2$.

## IV. β-SPECTRUM CALCULATIONS

Theoretical calculations of the two first forbidden nonunique transitions of $^{176}$Lu have been performed within the framework of the formalism described by Behrens and Bu¨hring in [77] based on Fermi theory. The general expression of the β-spectrum in this low-energy effective theory of the weak interaction is given, in relativistic units (~ = $m_e$ = $c$ = 1), by

$$\frac{dN}{dW} = \frac{G_\beta^2}{2\pi^3} F(Z,W) pW (W_0 - W)^2 C(W) X(W) \quad (10)$$

where $G_\beta$ is the Fermi coupling constant; $W$ is the total β-particle energy and $p$ its momentum; $W_0$ is the maximum available total energy; $pW(W_0-W)^2$ is the statistical shape that comes from the sharing of the momentum between the emitted leptons; and $F(Z,W)$ is the Fermi function that takes into account the Coulomb interaction between the β-particle and the daughter nucleus.

The shape factor $C(W)$ is a convolution of the nuclear structure and the lepton dynamics determined from a multipole expansion of the hadron and lepton currents, expressed as

$$C(W) = \sum_{K,k_e,k_\nu} \lambda_{k_e} \left[ M_K^2(k_e,k_\nu) + m_K^2(k_e,k_\nu) - \frac{2\mu_{k_e}\gamma_{k_e}}{k_e W} M_K(k_e,k_\nu) m_K(k_e,k_\nu) \right]. \quad (11)$$

Quantities labeled by the lepton quantum numbers $k_e$ and $k_\nu$ depend on the relativistic wave functions, with in particular $\lambda_1$ = 1. The main multipole order $K$ comes from the expansion of the nuclear current. We followed [78] for the calculation of $C(W)$, considering the dominant terms with $K$ = 1,2 and $k_e + k_\nu$ = 2,3. The lepton wave functions are expanded in powers of ($m_eR$), ($WR$) and ($\alpha Z$), with $R$ the nuclear radius and $\alpha$ the fine structure constant. This procedure avoids the calculation of overlaps between nuclear and lepton wave functions and the nuclear matrix elements, also called form factor coefficients, become independent of the lepton momenta. All the formulas used in the present work, assuming impulse approximation, are well described in [78].

The factor $X(W)$ stands for some additional corrections. The first one is for the atomic screening effect. The Fermi function and the $\lambda_{k_e}$ parameters directly depend on the electron wave functions, which have been determined as described in [13], i.e. considering the Coulomb potential generated by a uniformly charged sphere. The simplicity of such a potential makes possible the expansion of the lepton wave functions as described previously, but it prevents us including directly any screened potential. In a previous work [79], we developed a dedicated code for a full numerical calculation of the electron wave functions taking into account such screened potentials. In the present work, we have tabulated beforehand screened-to-unscreened ratios of $F(Z,W)$ and $\lambda_{k_e}$ at the required energies and corrected these quantities to account for screening in the β-spectrum.

The second correction corresponds to the atomic exchange effect. Correction of this effect for $k_e$ = 1 has been determined as described in [79] for the atomic $s_{1/2}$ orbitals and as complemented in [80] for the $p_{1/2}$ orbitals. This effect has recently been extended to the forbidden unique transitions as briefly described in [81] and we have used this formulation to determine the correction for $k_e$ = 2. Again, the correction factors have been tabulated beforehand at the required energies and applied during the β-spectrum calculation. Finally, radiative corrections are also included in $X(W)$. They are calculated as described in [82] and as a benchmark, we obtained excellent agreement with those determined in the survey of superallowed β-transitions in [83].

A realistic description of a nuclear state can be achieved via configuration mixing. Its many-particle wave function then results from a linear combination of single-particle (nucleon) wave functions. In β-decay, the transition amplitude between the initial and final nuclear states is determined by evaluating the corresponding onebody spherical tensor operator $T_\lambda$. It can be expressed as a weighted sum of the single-particle transition amplitudes [84]:

$$\langle \xi_f J_f || T_\lambda || \xi_i J_i \rangle = \hat{\lambda}^{-1} \sum_{a,b} \langle a || T_\lambda || b \rangle \langle \xi_f J_f || \left[ c_a^\dagger \tilde{c}_b \right]_\lambda || \xi_i J_i \rangle \quad (12)$$

where $\hat{\lambda}$ is the tensor rank, h$a$||$T_\lambda$||$b$i is a single-particle matrix element describing a nucleon-nucleon transition, and $\langle \xi_f J_f || \left[ c_a^\dagger \tilde{c}_b \right]_\lambda || \xi_i J_i \rangle$ is the one-body transition density.

The list of nucleon-nucleon transitions and their onebody transition densities have been determined employing the shell model code NuShellX@MSU [85]. Above a doubly-magic $^{132}$Sn core, we selected the *jj56pn* valence space and used the recommended *khhe* effective interaction [86]. Proton valence space spans from $1g_{7/2}$ to $1h_{11/2}$, in which 21 particles have to be distributed for $^{176}$Lu and 22 for $^{176}$Hf. Neutron valence space spans $1h_{9/2}$ to $1i_{13/2}$, in which 23 particles have to be distributed for $^{176}$Lu and 22 for $^{176}$Hf. Such a high number of particles leads to a non-tractable number of possible configurations. We have thus limited the valence space by assuming the lowest orbitals are completely full, i.e. $1g_{7/2}$, $2d_{5/2}$ and $2d_{3/2}$ for protons and $1h_{9/2}$ and $2f_{7/2}$ for neutrons. Other orbitals have been let free. The onebody transition densities have been calculated for the dominant $K$ values and are given in Table VIII for the main transition.

We followed the method depicted in [87] to estimate the harmonic oscillator frequencies required for the calculation of the single-particle matrix elements. To this purpose, the experimental root mean square charge radii of $^{176}$Lu and $^{176}$Hf have been taken in [88] and the nucleon configurations provided by NuShellX have been employed. We obtained $(\hbar\omega)_n$ = 7.594 MeV for the initial neutrons and $(\hbar\omega)_p$ = 6.833 MeV for the final protons. The equivalent uniform charge radii have been deduced and used in the calculations: $R_i$(Lu) = 6.868 fm and $R_f$(Hf) = 6.808 fm.

The $\beta$-decay formalism is totally relativistic due to the small rest mass of the $\beta$-particle, leading to particle wave functions with small and large components. In addition, the weak interaction is described by a linear combination of vector and axial-vector components. This results in different matrix elements that can be either vector or axial-vector, non-relativistic when only large components of the nucleon wave functions are needed, or relativistic when small components are involved. In the expansion of $M_K(k_e,k_\nu)$ in [78], one can see that in the present case the relativistic vector nuclear matrix element $^VF_{101}$ appears, for which an accurate value is of importance.

Table VIII. One-body transition densities (OBTD) of the dominant multipole orders in the main $\beta$-transition of $^{176}$Lu decay, as given by NushellX. Coulomb displacement energy for each nucleon-nucleon transition is also given.

| neutron → proton | | | OBTD | $\Delta E_C^{(3)}$ (MeV) |
|---|---|---|---|---|
| $K=1$ | | | | |
| $3p_{3/2}$ | → | $3s_{1/2}$ | 0.01779 | 15.415 |
| $3p_{1/2}$ | → | $3s_{1/2}$ | 0.05171 | 15.415 |
| $1i_{13/2}$ | → | $1h_{11/2}$ | -0.83750 | 15.957 |
| $K=2$ | | | | |
| $2f_{5/2}$ | → | $3s_{1/2}$ | 0.01600 | 15.006 |
| $3p_{3/2}$ | → | $3s_{1/2}$ | 0.02385 | 14.631 |
| $1i_{13/2}$ | → | $1h_{11/2}$ | -0.00769 | 15.551 |

However, NushellX is a non-relativistic nuclear structure model. One could identify the large component of the nucleon wave function to the non-relativistic wave function and estimate the small component from the large one, but the inaccuracy of such an approach has been seen for decades (see e.g. [89]). Another approach is to assume the conserved vector current hypothesis (CVC) that comes from gauge invariance of the weak interaction. One can then relate $^VF_{101}$ to the non-relativistic form factor coefficient $^VF_{110}$ by [77]

$$^VF_{101} \sim -\frac{R}{\sqrt{3}}\left(W_0-(m_n-m_p)+\Delta E_C\right){}^VF_{110} \quad (13)$$

with $m_n$ and $m_p$ the neutron and proton rest masses and $\Delta E_C$ the Coulomb displacement energy. Considering a uniformly charged sphere, the latter quantity is expressedby the usual formula

$$\Delta E_C^{(1)} = \frac{6}{5}\frac{\alpha Z_f}{R_f} = 18.273 \text{ MeV}. \quad (14)$$

One can also go back to the derivation of this formula and establish another one that depends on the initial and final nuclear radii:

$$\Delta E_C^{(2)} = \frac{3}{5}\frac{\alpha}{R_f}Z_f(Z_f-1) - \frac{3}{5}\frac{\alpha}{R_i}Z_i(Z_i-1) = 23.460 \text{ MeV} \quad (15)$$

where we used $R_i$(Lu) and $R_f$(Hf) as given above. It is noteworthy that a constant $\Delta E_C$ for every nucleonnucleon transition is an approximation. The Coulomb displacement energy was demonstrated a long time ago to possibly be sensitive to the mismatch between the initial and final nucleon wave functions [90]. As described in [77], we have assumed that the single-particle potential difference is determined by the average of the Coulomb potential $V(r)$ only:

$$\Delta E_C^{(3)} = \frac{\int_0^\infty g_f V(r) g_i \left(r/R\right)^K r^2 \mathrm{d}r}{\int_0^\infty g_f g_i \left(r/R\right)^K r^2 \mathrm{d}r} \quad (16)$$

where $g_i$ and $g_f$ are the radial large components of the initial and final nucleon wave functions, respectively. The calculated values for the different nucleon-nucleon transitions of interest in this work are given in Table VIII.

We present in Fig. 14 the spectra of the main transition calculated with the three different methods for determining $\Delta E_C$. We have considered the free-nucleon value of $g_A$, i.e. $g_A^{\text{free}}$ = 1.2763(15) as the result of the mean of two recent precise measurements [91, 92]. These spectra are compared to the measurement from this work and to an allowed spectrum, for which $C(W)$ = 1. Indeed, first forbidden non-unique transitions are usually treated as allowed when the $\xi$-approximation is fulfilled, i.e. $\alpha Z/R \gg (W_0-1)$ [13]. If including nuclear structure is clearly important, we also see the great influence of the sole Coulomb displacement energy.

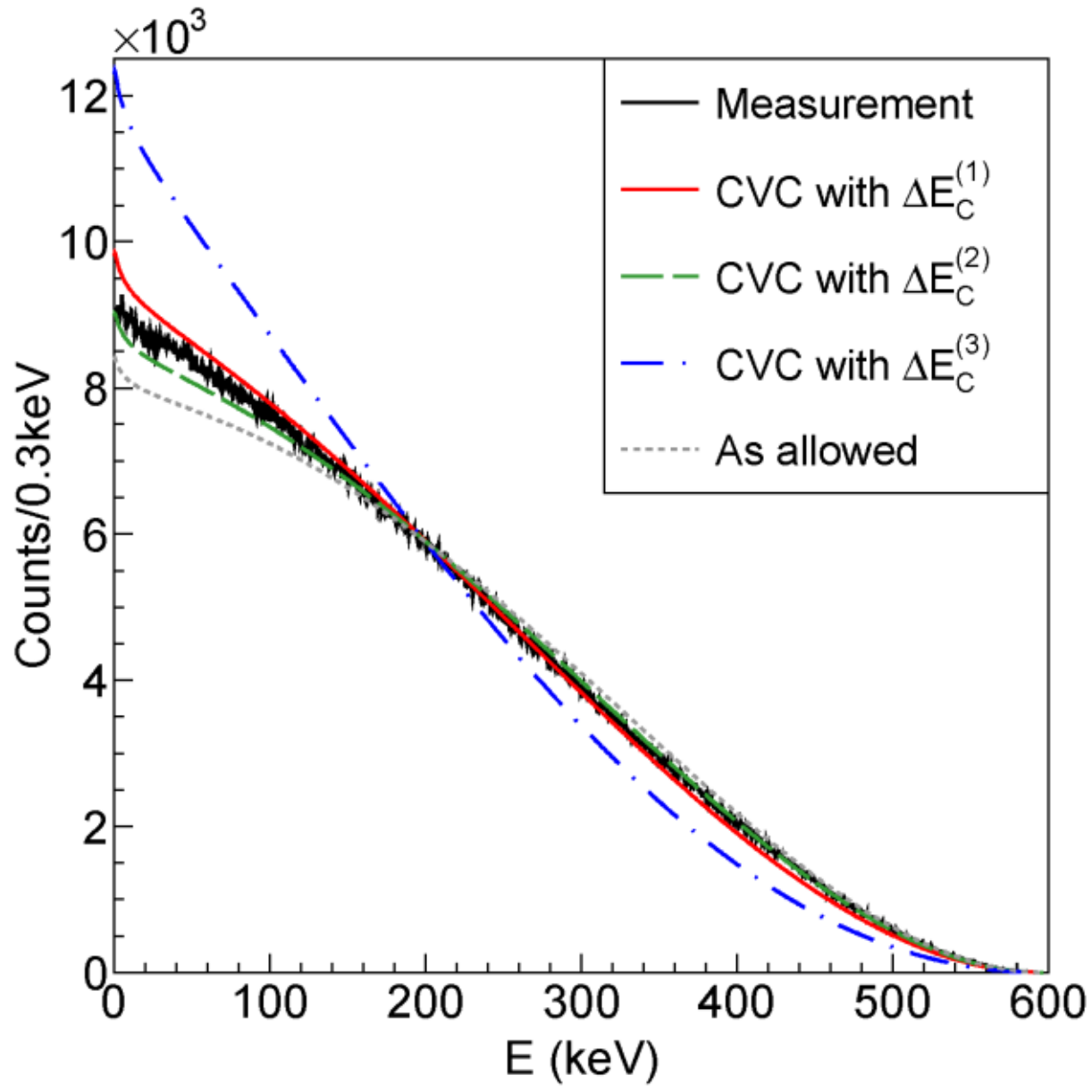

Figure 14. The measured $\beta$-spectrum of the main transition in $^{176}$Lu decay is compared to four different calculations. Theoretical spectra with nuclear structure have been computed using the CVC hypothesis and three different methods to determine the Coulomb displacement energy $\Delta E_C$. The freenucleon value of $g_A$ has been considered.

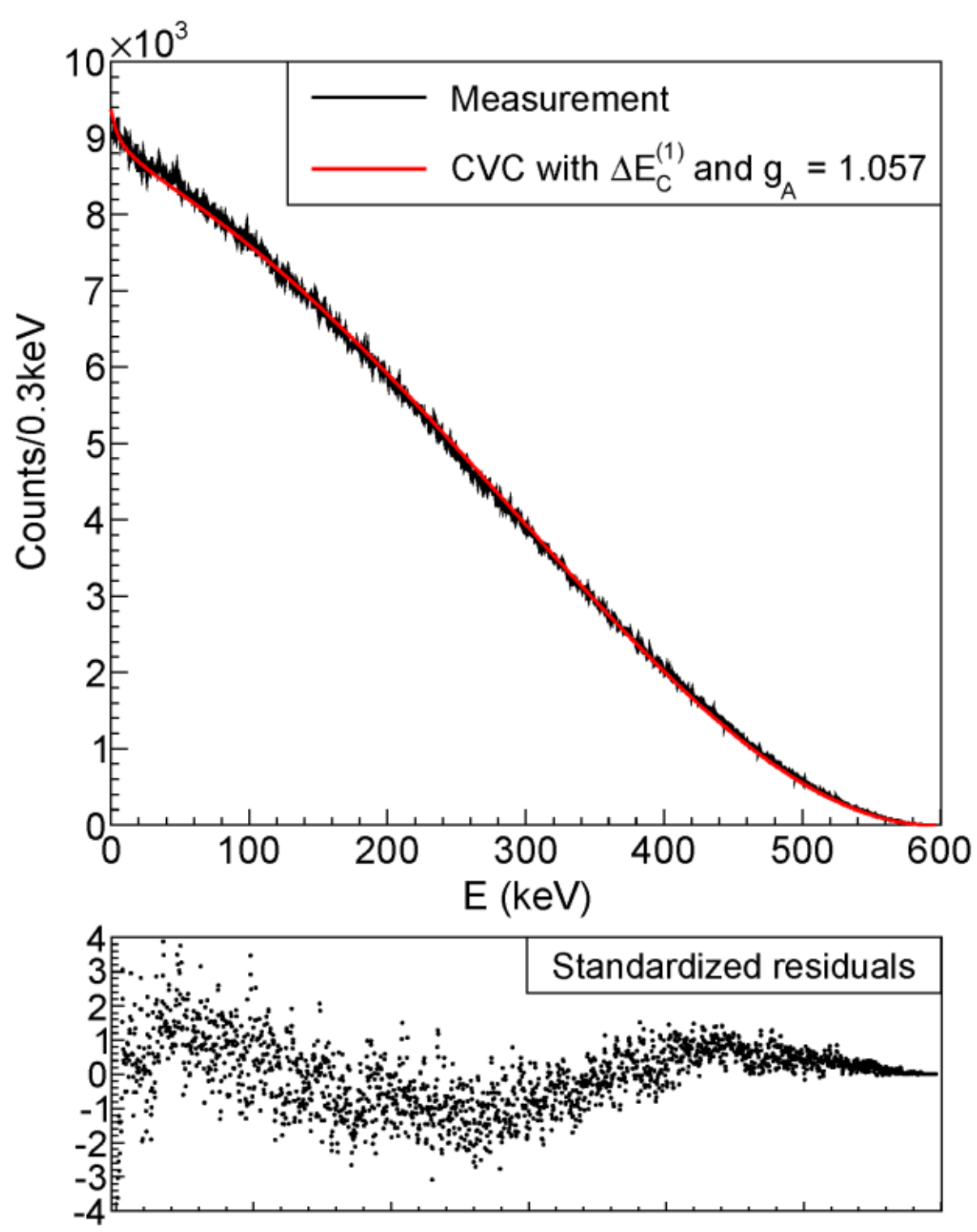

Figure 15. Theoretical spectrum of the main transition in $^{176}$Lu decay. An effective $g_A$ constant has been fitted to reproduce at best the measured spectrum. A non-linear trend clearly remains in the standardized residuals.

The value of the axial-vector coupling constant $g_A$ has been shown to potentially influence the spectrum shape of forbidden non-unique transitions, sometimes significantly [14, 93, 94]. Indeed, an adjustment may be necessary to reproduce some experimental observables because the nucleon-nucleon transition occurs in nuclear matter. A quenched value of $g_A$ can then take up a part of the mismodeling of nuclear structure, e.g. an approximate treatment of the many-nucleon correlations. Following a recent review [95], one can deduce a quenching factor for $^{176}$Lu decay from the quenching factor in infinite nuclear matter and estimate an effective value of $g_A^{\text{est}} = 1.1075$.

As some of our spectra with $g_A^{\text{free}}$ are not so far from the experimental spectrum, we have applied the "spectrum shape method" proposed in [96] to extract an effective $g_A$ value. We tried four possibilities: first, we kept $g_A^{\text{free}}$ and we adjusted the Coulomb displacement energy; next, $g_A$ was adjusted for the different $\Delta E_C$. Each possibility was found to give very similar agreement with the measurement and close standardized residuals. We present in Fig. 15 one of the results and the adjusted values are given in Table IX. Uncertainties on the parameters come from the fit procedure and also include the influence of the energy range considered. As expected, the value of $g_A$ strongly depends on the assumed Coulomb displacement energy used in CVC. The usual formula $\Delta E_C$ gives the closest effective value to $g_A^{\text{est}}$. It is noteworthy that a non-linear trend clearly remains in the standardized residuals, even if the latter lie within $\pm 2\sigma$.

The log$f$ values corresponding to each adjustment are also given in Table IX, where the uncertainties also includes the component due to the maximum energy. One can see that they are systematically negative but widely spread. The mean value is log$f$ = −0.91(14). From the partial half-life, evaluated from experimental results for the branching ratio and the total decay half-life [27], one can deduce for this transition log$ft$ = 17.17(14). This value is lower than in [27] where the transition was calculated as allowed with the LogFT code [64], leading to a log$f$ value of 1.093(2).

Finally, one has to mention that we have not been able to reproduce the measured spectrum of the second $\beta$-transition to the $8^+$ state of $^{176}$Hf. In addition, the spectrum shape has not been found to be sensitive to the $g_A$ value. The valence space in NushellX is probably too tightly constrained to obtain a realistic description of this nuclear state. As the calculations depend on the nuclear structure, the sensitivity of this $\beta$-transition to $g_A$ is not conclusive.

Table IX. Adjusted values that lead in each case to similar spectra and standardized residuals as shown in Fig. 15. The corresponding reduced-$\chi^2$ and log$f$ values are also given.

| $\Delta E_C$ (MeV) | $g_A$ | reduced-$\chi^2$ | $\log f$ |
|---|---|---|---|
| 20.527(46) | $g_A^{\rm free}$ | 1.278 | -0.835(19) |
| $\Delta E_C^{(1)}$ | 1.057(4) | 1.258 | -0.975(14) |
| $\Delta E_C^{(2)}$ | 1.560(5) | 1.296 | -0.679(12) |
| $\Delta E_C^{(3)}$ | 0.834(3) | 1.227 | -1.148(14) |

## V. DISCUSSION AND CONCLUSIONS

Coupled with present knowledge of scintillation processes, the relatively large natural activity in Lucontaining scintillators, has enabled precise observation of $\beta$-emissions of $^{176}$Lu and a much stricter formulation of the upper limit of its EC decay branches. Experimental shaping factors for the entire energy range of the dominant $\beta$-transition of $^{176}$Lu, with a statistics of $8.5 \times 10^6$ counts, have been made available. Moreover, even the least probable $\beta$-transition of $^{176}$Lu was measured in its entire energy range and shape factors formulated, albeit with a counting statistic limited to 83000 counts. As a result of efforts dedicated to observe EC decays, new upper limits were established for their branching ratios, confining their probability by almost 3 orders of magnitude.

The presently implemented self-scintillation method for $^{176}$Lu provided unmatched results, made robust by independent experimental measurements and theoretical evaluations. Nevertheless, room for further improvements exist. These include: the use of a coincidence spectrometer with enhanced energy resolution compared to NaI(Tl) for a more selective generation of gates; the use of a synchronized dual-channel signal-digitalization system allowing post-acquisition processing of coincidence; an improved pen-type PMT or an alternative scintillation light readout such as SiPM, for enhancing energy resolution of the $\beta$-spectroscopy and improving noise performances. Use of a custom geometry Lu-containing scintillator can also be considered. Small detector effects have been assessed empirically and the result allowed them to be neglected in the response function. Although satisfactory in the present experiments, this result might benefit from further investigations both experimentally and by means of simulations.

The Penning trap measurements performed in this work provide the first direct measurement of the $^{176}$Lu $\beta$-decay Q value. Along with the energies of the $6^+$ and $8^+$ daughter levels in $^{176}$Hf, the results provide precise end point energies for the $\beta$-spectra of the two $^{176}$Lu decay branches. For the dominant $\beta$-transition, the precision of the end point obtained by Penning trap measurements, i.e. 596.21(55) keV, validates the value obtained by self-scintillation of 596.6(9) keV. Similarly for the minor $\beta$-transition, the end point value of 195.30(55) keV obtained by Penning trap measurements validates that of 195.7(25) keV by self-scintillation.

Our theoretical studies of the main transition in $^{176}$Lu $\beta$-decay have led to very different effective $g_A$ values, and the residuals in Fig. 15 indicate that part of the shape is not well reproduced. A possible improvement of our modeling would be a more accurate treatment of the lepton current, e.g. with next-to-leading-order terms as in [96], that could make the spectrum shape less sensitive to the Coulomb displacement energy.

One might think about looking at partial half-lives, $t_{1/2}$, to select the adjustment that gives the best value. However, our study is based on a nuclear structure determined with a spherical shell model while $^{176}$Lu is well known to be strongly deformed. This leads to a hindered transition rate such as [97]

$$t_{1/2}^{\rm exp} = t_{1/2}^{\rm theo} / \left[F^{\Delta K - 1}\right]^2 \tag{17}$$

where $K$ is the appropriate quantum number corresponding to the projection of the total angular momentum on the symmetry axis ($\Delta K = 7$ in our case), and $F$ ' 0.15 is the reduction factor as determined in [97]. The four adjustments we have performed lead to $t^{\rm theo}{}_{1/2}$ values that can differ by up to a factor of three, but which are 13 orders of magnitude smaller than $t_{1/2}^{\rm exp}$. From these results, we deduced a reduction factor of $F = 0.0768(20)$. The adjusted $g_A$ value with $\Delta E_C$ gives the best corrected partial half-life but $g_A^{\rm free}$ with $\Delta E_C$ adjusted gives a very close value. A detailed analysis of $^{176}$Lu decay with a realistic nuclear structure that includes nucleus deformation is therefore required to extract a firm effective $g_A$ constant.

## ACKNOWLEDGMENTS

This research was supported by Michigan State University and the Facility for Rare Isotope Beams and the National Science Foundation under Contracts No. PHY1102511, PHY-1307233 and PHY-2111185. This material is based upon work supported by the US Department of Energy, Office of Science, Office of Nuclear Physics under Award No. DE-SC0015927 and DE-SC002538. The work leading to this publication has also been supported by a DAAD P.R.I.M.E. fellowship with funding from the German Federal Ministry of Education and Research and the People Programme (Marie Curie Actions) of the European Union's Seventh Framework Programme (FP7/2007/2013) under REA Grant Agreement No. 605728.

Part of the experimental work of the $\beta$-spectra measurement was performed as part of the EMPIR Project 15SIB10 MetroBeta. This project has received funding from the EMPIR program co-financed by the Participating States and from the European Union's Horizon 2020 research and innovation program.

The theoretical work was performed as part of the EMPIR Project 20FUN04 PrimA-LTD. This project has re EMPIR Project 20FUN04 PrimA-LTD. This project has received funding from the EMPIR program co-financed by the Participating States and from the European Union's Horizon 2020 research and innovation program.

---